\documentclass[%
 reprint,
 amsmath,amssymb,
 aps,nofootinbib,   
]{revtex4-1}
\usepackage{bm}
\PassOptionsToPackage{linktocpage}{hyperref}
\usepackage[hyperindex,breaklinks]{hyperref}
\usepackage{enumitem}
\usepackage{slashed}

\renewcommand{\theta}{\vartheta}

\newcommand{\bra}[1]{\ensuremath{\left< #1\,\right|}}
\newcommand{\ket}[1]{\ensuremath{\left|\, #1\right>}}

\usepackage{array}
\usepackage{mathtools}

\usepackage{etoolbox}

\begin{document} 

\title{Strong-$CP$ with and without gravity}

\author{Gia Dvali} 

\affiliation{
	Arnold Sommerfeld Center, 
	Ludwig Maximilians University, 	
	Theresienstra{\ss}e 37,
	 80333 Munich, Germany  
}
\affiliation{
	Max Planck Institute for Physics, 
	 F\"ohringer Ring 6, 80805 Munich, Germany
}


\begin{abstract}

Conventionally, the strong-$CP$ problem is assumed to be a naturalness puzzle, with the axion solution sometimes viewed as an ad hoc fix.
Gravity is either ignored or taken as a threat for the global Peccei-Quinn symmetry.  
We explain that the situation is fundamentally different.  
In gravity, axion is a matter of consistency imposed by the $S$-matrix:  Each gauge sector
must include axion with exact relaxation of the corresponding $\bar{\theta}$.
We show that this favors an 
alternative and remarkably simple formulation of the axion,
fully fixed by the gauge redundancy of QCD, without 
 involvement of a global symmetry.  
The axion mechanism is a Higgs effect for the QCD $3$-form, 
ensuring that physics is independent of $\bar{\theta}$
to all orders in operator
expansion.  A near-future experimental  detection of the neutron EDM will be an unambiguous signal of $CP$-violating physics beyond the Standard Model.  The axion coupling is tied to the scale of gravity.

 \end{abstract}


\maketitle

 \subsection{The message} 
 
 In the standard discussion of the strong-$CP$ puzzle,
gravity plays no useful role. The puzzle is formulated as a naturalness problem of the following 
essence.  
QCD has a continuum of vacua \cite{Tvacua1, Tvacua2}, conventionally labelled 
by the $CP$-violating vacuum angle $\theta$.  
These vacua belong to different superselection sectors. 
In fact, the physically measurable parameter is 
the quantity $\bar{\theta} \equiv \theta  + {\rm arg.det M_q}$,
where ${\rm arg.det. M_q}$ is the phase of the determinant 
of the quark mass matrix. 

  In quantum theory, $\bar{\theta}$ induces the electric dipole moment of 
  neutron (EDMN) \cite{Baluni:1978rf, Crewther:1979pi}. 
 The comparison of the resulting theoretical value with the current experimental 
  limit,  $|d_n| <  2.9 \times 10^{-26} e$cm \cite{Baker:2006ts},
  gives the bound, 
  \begin{equation} \label{BBartheta}
  |\bar{\theta}|  \lesssim 10^{-9}\,. 
  \end{equation} 
 Notice that there exists an additional contribution
 to EDMN, coming from the 
 breaking of $CP$-symmetry by the weak interaction  
 \cite{Ellis:1976fn, Shabalin:1979gh, Ellis:1978hq}. 
 However, this correction is too small for affecting the bound (\ref{BBartheta}).

Thus, the observations indicate that we live in 
a sector with a minuscule or zero $\bar{\theta}$. 
 This is puzzling.  
 
  Axion \cite{axion1, axion2} eliminates this vacuum structure by making $\bar{\theta}$ dynamical and by relaxing it to a $CP$-invariant ground state.  In the original model by Peccei and Quinn \cite{PQ}, the axion emerges as a pseudo-Nambu-Goldstone boson 
 of an anomalous global symmetry  $U(1)_{PQ}$.   
  However, 
   this scenario is subjected to the following potential criticism. 
   
  First, it appears rather artificial, since axion is 
  designated to do a single job. Its introduction 
  is not justified by any visible consistency requirement. 
  
  Secondly, one may argue that the original 
  puzzle is replaced by a new one, since the phenomenological 
  bound (\ref{BBartheta}) 
  requires an extraordinary precision of the global
  $U(1)_{PQ}$-symmetry, modulo the 
  chiral anomaly of QCD. 
  This is a mystery, since 
 the global symmetries are not protected by any known 
 fundamental principle. The puzzle is sharpened by the fact that the
 axion mechanism is based on the explicit breaking of this very symmetry by the QCD anomaly.
 It is then totally unclear what prevents the explicit breaking of 
 $U(1)_{PQ}$ by other sources. 
 
 Thirdly, the theory suffers from the lack of predictivity.  
 Due to the arbitrariness of the explicit breaking of $U(1)_{PQ}$-symmetry,  the quantity $\bar{\theta}$ is uncalculable. 
  Correspondingly, the EDMN cannot be predicted.  
    
   In the above picture gravity is either ignored or is viewed as a potential threat. This is because, at the level of an effective field theory  (EFT), gravity has no obligation to respect global symmetries. 
 Due to this, it is usually assumed that 
 $U(1)_{PQ}$ is broken by the high-dimensional 
 operators of gravitational origin. If so, they can jeopardise 
 the axion mechanism.      

 The goal of the present paper is to 
 argue that the situation is fundamentally different.
   
  First, in  
 gravity the absence of the strong-$CP$-violating $\theta$-vacua 
 is a consistency requirement. 
 The most transparent reason is the $S$-matrix formulation of gravity \cite{Dvali:2020etd}. 
This formulation is  incompatible with the existence 
of a non-degenerate landscape of $CP$-violating $\theta$-vacua. 
This demands the existence of axions that relax the 
landscape towards the $S$-matrix vacua \cite{Dvali:2018dce}.  
  As we shall show, the relaxation must be exact, persistent to all orders in  the operator expansion. 
 
 Thus, at first glance, gravity provides the two contradictory 
 messages. On one hand, it demands the exactness of the axion mechanism.  On the other hand, it exhibits no respect for global  symmetries, such as $U(1)_{PQ}$.  
 
 Within the standard Peccei-Quinn formulation of axion, 
 the reconciliation of these two tendencies is difficult to 
 understand.

  In the present work, we show that this is accomplished by an alternative theory of axion, as previously discussed in 
  \cite{Dvali:2005an}. In this formulation, there exist no global symmetry.  Instead, the axion, introduced as  a $2$-form
 $B_{\mu\nu}$, shares the gauge redundancy of gluons.  With no 
further assumptions, the gauge redundancy
fixes the structure of the theory making
it fully insensitive to deformations caused by arbitrary operators.

In this theory,  
the axion mechanism   
  is described as a Higgs-like effect
 for the Chern-Simons   $3$-form of QCD.
 Eating up the $B_{\mu\nu}$-axion,  the 
 $3$-form becomes massive.  Correspondingly, the 
 topological susceptibility of the vacuum (TSV) vanishes and 
 $\bar{\theta}$ becomes unphysical.

 Due to gauge protection, the axion vacuum remains exactly $CP$-invariant  under any continuous deformation of the theory. 
 It is insensitive towards arbitrary sort of UV-physics \cite{Dvali:2005an}.
  We show that while the gravitational effects, such as the virtual black holes, 
 may contribute into the masses of axion and $\eta'$-meson, they 
 are unable to destabilize the axion mechanism.  
  This guarantees the fulfilment of the constraint imposed by the gravitational $S$-matrix to all orders. 
  No such protection exist in the ordinary Peccei-Quinn
 realization in which the axion represents a Goldstone phase of a complex scalar. 
 
 We explain this difference carefully.  As shown in \cite{Dvali:2005an}, the $B_{\mu\nu}$ formulation 
can be dualized to a theory of a pseudo-scalar axion 
$a$ with arbitrary shape of the potential. 
 However, the caveat is that in the dualized  theory the axion enters with an integration constant. This constant encodes the information about the 
 gauge structure of the $B_{\mu\nu}$-theory.  
 
 Such an integration constant is absent in a generic EFT of a speudo-Goldstone axion with arbitrary explicit breaking of $U(1)_{PQ}$. 
 This is the key for understanding of how the
 protection by the gauge redundancy of the $B_{\mu\nu}$-theory is  
 transferred to its exact duals.  
  At the same time, this explains the lack of  protection  in the ordinary Peccei-Quinn case.

In the alternative theory, the axion emerges as an inseparable
part of the  gauge redundancy of QCD.  This redundancy is the 
fundamental reason for the elimination of $\theta$-vacua. 
However, at the level of EFT, the 
axion decay constant $f_a$ enters as a free parameter.  
 Although this formulation 
is independent of gravity, it is strongly motivated by it. 
  Therefore, the painted picture hints towards 
 the gravitational origin of the scale $f_a$. 
 
 The exactness of the axion mechanism 
 has important phenomenological consequences.  
 First, it endows the theory with the predictive power. 
  Due to vanishing of $\bar{\theta}$, the EDMN becomes 
 calculable in terms of other parameters.  
  For example, within the minimal setup,  a sole source of  EDMN is predicted to be the breaking 
of $CP$-invariance by the weak interaction. 
This contribution is calculable
\cite{Ellis:1976fn, Shabalin:1979gh, Ellis:1978hq} and 
is well below the sensitivity 
of current and prospective experiments (see, e.g., \cite{n2EDM:2021yah}).  

 This may sound disappointing but it is not. As a flip side of the coin,
 the EDMN becomes a direct probe for new physics. 
An experimental detection of EDMN will be an unambiguous sign of a $CP$-violating physics coming from beyond the Standard Model.

 The discussion below is structured as follows. 
 We first discuss the essence of the strong-$CP$ puzzle in 
 the absence of gravity.   We do this in the language of 
 \cite{Dvali:2005an},  which is most suitable for preparing the stage for gravity.  Next, we discuss the role of gravity 
 with various implications and conclude. 
 In particular,  we compare the effects of gravity on the 
 Peccei-Quinn formulation of QCD axion and on alternative $B_{\mu\nu}$-formulation. 
 
    Although, it is assumed that the reader is familiar with the standard 
 description of the strong-$CP$ problem and of its  Peccei-Quinn solution, we try to make the presentation self-contained. 
            
    We note that, although the exactness of the axion mechanism 
  is justified by the $S$-matrix formulation of gravity,  most of 
  our analysis and conclusions are independent of it. 
  For example, the  $3$-form formulation of the axion mechanism 
  and its protection by the gauge redundancy \cite{Dvali:2005an} are universal.  Likewise, the subsequent phenomenological implications for EDMN are valid regardless of gravity. 
  
   Note that throughout the paper all irrelevant numerical factors 
   will be either set to one or rescaled in redefinitions of fields
   and parameters with each step.  In this way we avoid
   carrying them around.  
    The scale of the $3$-form theory, such as 
   the QCD scale, unless shown explicitly, will also be set to one.  
 
  \subsection{$3$-form vacua}  
 
  The deep meaning of the strong-$CP$ problem 
  and of its solution by axion is especially transparent in the language of   the $3$-form gauge theory 
  \cite{Dvali:2005an}. 
 The advantage of this formulation is that it relies on basic properties of QFT, such as, the power of gauge redundancy and the spectral representation of correlators.  
  Following  \cite{Dvali:2005an}, we first
   discuss the effect for a generic $3$-form and later 
   focus on gauge theories such as QCD.

   A  massless $3$-form field $C_{\alpha\beta\gamma}$ propagates no 
   degrees of freedom. The Lagrangian can be chosen 
 as a generic algebraic function (${\mathcal K}$) of the gauge-invariant 
 field strength $E  \equiv \epsilon^{\alpha\beta\gamma\delta}\partial_{\mu}C_{\alpha\beta\gamma}$,  
  \begin{equation} \label{LofC} 
  L = {\mathcal K}(E)\,. 
 \end{equation} 
For example, in the simplest case we can take
 ${\mathcal K}(E) \, = \, E^2$.
 In the absence of sources, there is no restriction on the 
 ${\mathcal K}$-function.  In the presence of sources 
 that satisfy a quantization condition, this
 function is restricted accordingly.  
 
 The massless $3$-form exhibits the following gauge redundancy, 
 \begin{equation} \label{gaugeC} 
 C_{\alpha\beta\gamma}(x) \rightarrow 
 C_{\alpha\beta\gamma}(x) +  \partial_{[\alpha}\Omega_{\beta\gamma]}(x)\,,
 \end{equation} 
where $\Omega_{\beta\gamma}(x)$ is an antisymmetric 
$2$-form function.

  The equation of motion, 
    \begin{equation} \label{Evac}
 \partial_{\mu} \frac{\partial {\mathcal K}(E)}{\partial E} \, = \, 0\,,
 \end{equation} 
  is  solved by an 
 arbitrary constant, 
    \begin{equation} \label{Ezero}
 E \, = \, E_0 \, = \, {\rm constant}\,. 
 \end{equation} 
 This represents a $3+1$ dimensional 
 analog of the electric field in $1+1$ dimensional 
 Schwinger model.  
 The constant ``electric 
 field"   $E_0$  is also analogous  to an ordinary Maxwell electric field created by 
 a charged capacitor plate located at the boundary.  
 Of course, in case of the $3$-form, the field $E_0$ is a Lorentz 
pseudo-scalar, with no preferred direction.  
Due to these analogies, one can say that a massless $3$-form is in the ``Coulomb" phase.

  The constant electric field contributes into the vacuum energy
  density $\sim E_0^2$. 
  Thus, a massless $3$-form creates a continuum landscape 
  of  vacua with  different values of the vacuum energy density. 
  These vacua define the distinct super-selection sectors.
  No transitions between them are possible.

The presence of dynamical sources allows  
some jumps. 
By gauge redundancy (\ref{gaugeC}), the source must be conserved. The examples are the $2$-branes or the
axionic domain walls.  The part of the action describing 
a gauge invariant coupling of $C$ to such a source is 
 \begin{equation}\label{BraneCoupling} 
Q \int dX^{\alpha} \wedge dX^{\beta} \wedge 
 dX^{\gamma}  C_{\alpha\beta\gamma} 
  \equiv  Q \int_{2+1} C \,,     
 \end{equation}
where $Q$ is the charge and the integration is performed over the world volume of a $2$-brane with the embedding 
coordinates $X^{\mu}$. 

 It is easy to see that  across 
 the source, $E$ experiences a jump determined by $Q$. 
 For example, for a static brane
coincident with $z=0$ plane, the change is 
$\Delta E \equiv  E(z>0) - E(z<0) = Q$.  

 In general,  the sources can obey the
quantization conditions, {\it a la} Dirac.  This 
puts certain restrictions on the kinetic function ${\mathcal K}(E)$.

 In the presence of sources, some discrete transitions 
 among the vacua are possible.  This splits the vacuum landscape into families of vacua that can be connected via quantum tunnelling.  However, 
 as long as the $3$-form is massless, the vacua are not degenerate and a continuum of superselection sectors is maintained.

 In summary, a massless $3$-form creates a continuous 
landscape of non-degenerate vacua.  These vacua are physically distinct as they differ by the VEV of a gauge-invariant order parameter $E_0$.  
This quantity is parity-odd. Correspondingly, it 
breaks both $P$ and $CP$.  
Due to this, different vacua have different strengths
of $CP$-violation.  This is a physically measurable effect.  

Notice that 
breaking of $CP$ comes from the VEV and therefore 
 is spontaneous.  This is important because a spontaneous  
$CP$-violation cannot be eliminated by imposing a 
symmetry on the  Lagrangian. 
 Indeed, imposing the exact parity symmetry implies 
 that ${\mathcal K}(E)$ is an even function of $E$. 
 For example, ${\mathcal K}(E) = \frac{1}{2}E^2$.
 Nevertheless, the vacuum equation (\ref{Evac}) is solved 
 by an arbitrary non-zero VEV $E_0$. 
 All solutions are the legitimate vacua. 
 One can of course demand that $E_0 \, = \, 0$.
 However, this is merely a choice of the vacuum, 
 not more justified than any other choice.

\subsection{Higgsing of $3$-form by axion} 

 There exists an unique way for eliminating the vacua with $E_0 \neq 0$.  For this it is necessary to  make 
 the $3$-form massive.   
 That is, the $3$-forms must be in Proca (or Higgs) phase. 
 A massive $3$-form propagates one degree of freedom and 
 is equivalent to a massive pseudoscalar. 
 The VEV of the corresponding electric field $E_0$, is strictly zero.     
 
   The only known gauge invariant (and therefore 
   consistent) way of endowing the 
 $3$-form with the mass, is via coupling it to an axion.
 At the level of the low energy EFT, 
 the axion can be introduced in two ways. 
 We discuss the formulation in terms of a pseudoscalar field,  
 $a(x)$,  first. 
 
    As shown in  \cite{Dvali:2005an}, this theory is equivalent to a theory of a massive axion with the potential that depends on an integration constant.  This integration constant is crucial for  keeping in tact the global 
 structure of the theory.  

    The form of the potential is determined by the 
 kinetic function of  the $3$-form. 
 The Lagrangian describing the generation of 
 mass for a $3$-form is, 
 \begin{eqnarray} \label{LaxionK}
 L &=&  {\mathcal K}(E) +  {1 \over 2} (\partial_{\mu} a)^2 -  {a \over f_a} E \,,
\end{eqnarray}
where $f_a$ is an axion decay constant. 
Notice that the axion shift by an arbitrary constant 
$\alpha$, 
\begin{equation}\label{axionshift}  
\frac{a}{f_a} \rightarrow \frac{a}{f_a} + \alpha \,, 
\end{equation} 
results in a total derivative in the Langangian. This can be written
 as the shift of the action by a boundary term, 
\begin{equation}\label{axionshift}  
\delta S = \int_{3+1} \, \alpha \,E \, = \,  \alpha \int_{2+1} C \,, 
\end{equation} 
where the last integration is taken over the boundary.  
 For generating a mass of $C$, it is crucial that 
 this global shift symmetry is not broken by any additional 
 source (see later).   
 
  The equations of motion are,  
\begin{eqnarray} \label{EQE} 
\partial_{\mu} \frac{\partial {\mathcal K}(E)}{\partial E}
  &=&  \partial_{\mu} \frac{a}{f_a}\,, \\
  \label{EQE2} 
 \Box a +  \frac{1}{f_a}E& = &0  \,.      
\end{eqnarray}
Integrating the first equation, we get, 
  \begin{equation}\label{KVa}
  \frac{\partial  {\mathcal K}(E)}{\partial E} 
   = \frac{a - a_0}{f_a} \,,
 \end{equation}
where $a_0$ is an arbitrary integration constant. 
This is an algebraic equation which expresses 
$E$ as a function of the r.h.s. of (\ref{KVa}).
Thus, we can write,  
\begin{eqnarray}\label{VK1}
E(z) = {\bf inv} \frac{\partial  {\mathcal K(z)}}{\partial z}   \,,      
\end{eqnarray}
where ${\bf inv}$ stands for the inverse-function and   
$z \equiv (a-a_0)/f_a$. 
From the equation (\ref{EQE2}), it is clear that 
$E(z)$ at the same time represents the first derivative of the axion 
potential with respect to axion, 
\begin{eqnarray}\label{VK2}
E(z) \, = \, \frac{\partial V(z)}{\partial z} \,.     
\end{eqnarray}
    This allows to rewrite the axion  
     equation in a conventional 
   form, 
\begin{eqnarray}\label{THa}
 \Box a + \frac{\partial V(a-a_0)}{\partial a} & = &0  \,.      
\end{eqnarray}
The theory (\ref{LaxionK}) is thus equivalent 
to a theory of a pseudoscalar with the following Lagrangian, 
 \begin{eqnarray} \label{LaxionA}
 L_a &=& {1 \over 2} (\partial_{\mu} a)^2 -  V\left (\frac{a-a_0}{f_a} \right ) \,.
\end{eqnarray}
The potential is determined from the equations
(\ref{VK1}) and  (\ref{VK2}) which for convenience we combine 
in one equation
\begin{eqnarray}\label{VK}
\frac{\partial V(z)}{\partial z} = {\bf inv} \frac{\partial  {\mathcal K}(z)}{\partial z}  = E(z)  \,.      
\end{eqnarray}

Very important:  The equation (\ref{VK2}) guarantees that $E$ vanishes at any extremum of the axion potential for any 
choice of $a_0$. 
 Thus, the would-be  
superselection sectors disappear and there is an unique
vacuum with $E=0$.  Notice, a discrete degeneracy among the 
vacua is still possible, but all of them have $E=0$ and 
therefore conserve $CP$.

As emphasized in \cite{Dvali:2005an},
the dependence of $V(a-a_0)$ on the integration constant $a_0$
is crucial for carrying along the information about the 
invariance of the original theory (\ref{LaxionK}) with respect 
to the axion shift symmetry  (\ref{axionshift}). 

 In other words, the theory of a massive $3$-form represents  a theory of a pseudo-scalar with a special global structure of the vacuum. This structure is encoded in the integration constant $a_0$.  
 
 This global structure is not present for the generic 
 pseudo-scalars.  For example,  the pseudo-Goldstone bosons
obtained by introduction 
of arbitrary explicit breaking operators do not possess the 
integration constants.  
  Correspondingly, such explicit-breaking operators cannot emerge from the couplings to $3$-forms. 
  
 As we show later,  this difference has important implications for 
 the UV-completion of the theory of axion and for its  dual formulation.

 \subsubsection{Examples}

Let us demonstrate the above on examples 
with particular forms of the $3$-form kinetic function.

  For the simplest choice,  
${\mathcal K}(E) = \frac{1}{2} E^2$,   
the axion potential has the form, 
$V(a) = \frac{m_a^2}{2} (a-a_0)^2$, with $m_a^2 = f_a^{-2}$. 
The electric field is given by $E = \frac{a-a_0}{f_a}$.
 The vacuum is at $a =a_0$, which  implies 
 $E =0$.  
 
 Another example  discussed in \cite{Dvali:2005an} is, 
 \begin{equation} \label{Ecos}
 {\mathcal K}(E) = E \arcsin(E)
 + \sqrt{1-E^2} \,.
 \end{equation}
 This form is important as it generates a periodic axion potential of the form, 
 \begin{equation} \label{Vcos}
  V(a) \, = \, - \,  \cos\left (\frac{a-a_0}{f_a} \right) \,.
 \end{equation}
 Again, in both cases, the presence of the integration constant 
 $a_0$
 is important for embedding such pseudo-scalars in a 
 theory (\ref{LaxionK}) 
 with a $3$-form and with the shift symmetry (\ref{axionshift}).

\subsection{Dual formulation: Axion as gauge theory} 
 
 As shown in \cite{Dvali:2005an}, 
  for a theory of axion (\ref{LaxionA}) with 
  an arbitrary potential $V(a-a_0)$, there exists an exact dual formulation, in which 
axion $a$ is replaced by an antisymmetric Kalb-Ramond
$2$-form $B_{\mu\nu}$. The Lagrangian is, 
 \begin{eqnarray} \label{dual} 
 L \, &=& \, {\mathcal K}(E) \, + \, m_a^2(C - dB)^2 \,,
\end{eqnarray}
where $dB \equiv \partial_{[\alpha}B_{\mu\nu]}$. 
This is equivalent to a theory of 
axion with the equation of motion (\ref{THa}), where 
the potential $V(a-a_0)$ is determined from 
the function ${\mathcal K}(E)$ via (\ref{VK}).  
 
This theory exhibits a gauge redundancy under which 
$C$ and $B$ transform as,   
 \begin{eqnarray} \label{gaugeBC} 
 C_{\alpha\beta\gamma} &&\rightarrow 
 C_{\alpha\beta\gamma} +  \partial_{[\alpha}\Omega_{\beta\gamma]}\,, \nonumber \\
 B_{\alpha\beta} &&\rightarrow 
 B_{\alpha\beta} +  \Omega_{\alpha\beta}\,. 
 \end{eqnarray} 
 
  On top of this, the theory exhibits an additional 
 gauge redundancy, acting solely on $B_{\mu\nu}$, 
  \begin{equation} \label{Bextra}
B_{\mu\nu} \, \rightarrow \, B_{\mu\nu} \, + \, \partial_{[\mu}\xi_{\nu]} \,,
\end{equation}
 where $\xi_{\nu}(x)$ is an arbitrary one-form. 
 Due to this gauge redundancy,  $B_{\mu\nu}$ propagates 
 a singe degree  of freedom. 
 
 As explained in  \cite{Dvali:2005an}, the gauge symmetry ensures the protection of axion mechanism against arbitrary UV-physics. In particular, 
 no continuous  deformation of the theory can un-Higgs the 
$3$-form.  

 This may sound rather puzzling, since it appears  
 that in dual theory  (\ref{LaxionK}) the 
 continuous deformation has a different effect. 
 For example, it is easy to show that an explicit breaking of the axion shift symmetry (\ref{axionshift}) by an arbitrary additional potential
 $V_{expl}(a)$, renders the 
 $3$-form massless. 
 
  The key to the puzzle, is in paying attention 
  to the integration constant.  
  A continuously deformed $a$-theory, without an accompanying 
  integration constant $a_0$, cannot be dualized into a 
  gauge invariant theory of $B_{\mu\nu}$. 
   That is, the only deformations of the axion potential 
  permitted by duality are the deformations 
   by the functions of $a-a_0$. Such deformations leave the 
   $3$-form in the Higgs phase. Correspondingly, they have 
   gauge duals.

 In order to see this,  we perform the 
 dualization  as  its done in \cite{Dvali:2005an}. 
  As the first step, we treat the field strength of $B$  as a fundamental 
$3$-form,   $X_{\alpha\mu\nu} \equiv \partial_{[\alpha}B_{\mu\nu]}$, 
and impose the constraint, 
\begin{equation} \label{dX}
 \tilde{dX} \equiv \epsilon^{\alpha\beta\mu\nu}\partial_{\alpha}X_{\beta\mu\nu} = 
 0\,, 
\end{equation} 
via a Lagrange multiplier field $a(x)$.  The Lagrangian thus becomes, 
\begin{eqnarray} \label{dualX} 
 L \, &=&\,  {\mathcal K}(E)\, + \, m_a^2(C - X)^2 \,  +\,  
 a \tilde{dX}\,.
\end{eqnarray}
Integrating out $X$ through its equation of motion, 
we arrive to the theory (\ref{LaxionK}), which describes 
the axion $a$ coupled to the $3$-form  $C$. 

 As the next step, we 
integrate out $C$ through the equation of motion 
(\ref{EQE}). As we have already seen, this gives an effective theory 
of axion (\ref{LaxionA}) with the potential 
$V(a-a_0)$ determined through the equation
(\ref{VK}).  

The above two-step process shows that the theory of $B_{\mu\nu}$
is dual to a theory of axion $a$ with the potential $V(a-a_0)$
that depends on $a-a_0$.  The integration constant $a_0$ carries  the information about 
the shift symmetry (\ref{axionshift}) of (\ref{LaxionK})
and the gauge symmetry of $B_{\mu\nu}$-formulation.   
In this way, the global structure of the theory of $a$-axion, obtained via the dualization from the gauge theory of  $B_{\mu\nu}$, differs from 
a generic theory of axion with the potential $V(a)$.

  To conclude this section, let us summarize.
  A massless  $3$-form creates the 
 superselection  sectors 
  parameterised 
 by the VEV of its field strength 
 $E_0$.  The vacua have different energies. 
 All except $E_0=0$ break parity and $CP$-symmetry. 
 The breaking 
 of symmetry is spontaneous, as it comes from the 
 VEV. 
  Due to this,  the vacua cannot be forbidden by imposing $P$, $CP$ or any other discrete symmetry on the Lagrangian.  
  As usual, the spontaneous breaking implies that 
 the vacuum 
  does not respect the symmetry of the Hamiltonian. 
 
  The only consistent way for eliminating the 
  $CP$-violating 
  superselection sectors, is 
  via making the $3$-form massive. 
  This requires an axion.

   The formulation (\ref{dual}) in terms of $B_{\mu\nu}$
  makes the analogy with the Higgs effect very transparent.  
  The  $3$-form gets a mass by ``eating up" the  
  $2$-form $B_{\mu\nu}$ which plays the role of the 
  St\"uckelberg field. 
  
  Upon dualization and integration out of the $3$-form, 
 we arrive to a theory of a pseudo-scalar axion $a$
 (\ref{LaxionA}) with 
 the potential $V(a-a_0)$ that is a function of $a-a_0$. 
  This global dependence 
  on the integration constant $a_0$ reflects the gauge redundancy of the 
 $B_{\mu\nu}$-theory (\ref{dual}) and the respective shift symmetry 
 (\ref{axionshift}) of the $a$-theory (\ref{LaxionK}). 
 The shape of the potential is determined  
by the kinetic function ${\mathcal K}(E)$.

\subsection{$3$-form description of $\theta$ vacua of gauge theories} 

  The superselection sectors (\ref{Ezero}) of a massless 
  $3$-form theory (\ref{LofC}) are of direct relevance     
 for the $\theta$-vacua of non-abelian gauge theories. 
  As discussed in \cite{Dvali:2005an}, 
  the $\theta$-vacua in $SU(N)$ gauge 
  theory, 
  represent a particular case of the $3$-form vacua. 
  Their structure is fully captured by the theory (\ref{LofC}).  
   
 In $SU(N)$ gauge theory, the massless $3$-form  $C$  of (\ref{LofC})
 originates from the Chern-Simons $3$-form, 
  \begin{equation} \label{CS}
   C_{\mu\nu\alpha}^{(\rm{CS})} \equiv {\rm tr} (A_{[\mu}\partial_{\nu}A_{\alpha]} + {2 \over 3}A_{[\mu}A_{\nu}A_{\alpha]}) \,,
   \end{equation}
   where $A_{\mu} \equiv A_{\mu}^bT^b$ is an $N \times N$ 
gluon matrix and  $T^b$ are the  generators of $SU(N)$
with $b=1,2,...,N^2-1$ the adjoint index.  The trace is taken over the color indexes.  Under the $SU(N)$ gauge transformation, 
$A_{\mu} \rightarrow U(x)A_{\mu} U^{\dagger}(x) 
+  U^{\dagger} \partial_{\mu} U$ with  
$U(x) \equiv e^{-i\omega(x)^bT^b}$, the $3$-form 
shifts as (\ref{gaugeC}) with 
$\Omega_{\mu\nu} = {\rm tr} A_{[\mu}\partial_{\nu]}\omega$. 
The gauge invariant field strength is $E^{(\rm{CS})} 
\equiv dC^{(\rm{CS})}$. 

In more traditional notations,  $E^{(\rm{CS})} \equiv F_{\mu\nu}\tilde{F}^{\mu\nu}$, where $F$ is the gluon field strength and $\tilde{F}^{\mu\nu} \equiv \epsilon^{\mu\nu\alpha\beta}F_{\alpha\beta}$ is its dual.  

The choice of a particular $\theta$-vacuum 
is parameterized by the introduction of the following 
term in the Lagrangian, 
 \begin{equation}\label{Boundary} 
 \int_{3+1} \theta E^{(\rm{CS})} = \theta \int_{2+1} C^{(\rm{CS})} \,.      
 \end{equation}
In the last expression the integral is taken over the world-volume 
coordinates of the boundary.   
This form shows that $\theta$ represents a charge 
of the boundary under $C^{(\rm{CS})}$.

 The correspondence between the $\theta$-vacuum and  
 the $3$-form vacuum goes through 
 the TSV, 
  \begin{eqnarray}
    \label{EEcorr}
\langle E^{(\rm{CS})},E^{(\rm{CS})}\rangle_{p\to 0} &&\equiv \\
\equiv \lim\limits_{p \to 0}\int d^4 x e^{ipx} \langle T [E(x)^{(\rm{CS})},E(0)^{(\rm{CS})}]\rangle &&=\mathrm{const} \nonumber\,,
\end{eqnarray}
where $T$ stands for time-ordering and  $p$ is a four-momentum. 
As it is well-known, the existence of the $\theta$-vacua is equivalent 
to a non-vanishing value of the above correlator.

The language of TSV is most convenient for our purposes,
since it is insensitive to particularities of physics that 
makes this correlator non-zero.  
In the original 
proposal by 't Hooft, these are instantons. However, 
our discussion is equally receptive to any other source
within QCD or beyond.

   The important thing is that the existence of 
 $CP$-violating $\theta$-vacua is equivalent to having 
 a non-zero TSV, and vice versa. 
  This information is sufficient for  
  reaching our conclusions. 
  The language of TSV makes is very transparent that
 the $\theta$-vacua represent the vacua with 
 different  VEVs of the Chern-Simons electric field $E^{(\rm{CS})}$. 
  
 In order to see this, let us first  notice that the expression (\ref{EEcorr}) 
  implies the existence of a pole at $p^2=0$ in the 
  correlator of the two Chern-Simons $3$-forms.
 That is, the  K\"all\'en-Lehmann 
   representation of  this correlator has the following form, 
    \begin{equation}
    \label{CCcorr}
\langle C^{(\rm{CS})},C^{(\rm{CS})}\rangle =  \frac{\rho(0)}{p^2} + \sum_{m\neq 0} 
\frac{\rho(m^2)}{p^2 - m^2} \,, 
  \end{equation} 
   where $\rho(m^2)$ is a spectral function.   
    The key point here is that 
  $\rho(0)$
  is non-zero.  Without loss of generality,  we set it to one. 
   The poles $m^2 \neq 0$ are separated from $m^2=0$ by finite gaps. This follows from the 
  fact that, in the absence 
  of massless quarks, the $SU(N)$ gluodynamics is a 
 theory with a mass gap. 
 
  The  entry with $m=0$ does not violate this rule. This is due to the fact that it corresponds to a massless $3$-form which contains 
  no propagating degrees of freedom. 
  Despite this feature, the $m=0$ component is very important, since it  
is the one responsible for the vacuum structure of QCD.  
   
   To summarize, from non-zero TSV (\ref{EEcorr}) it follows 
   that the Chern-Simons $3$-form operator $C^{(\rm{CS})}$
   contains a massless 
   $3$-form field $C$, plus the tower of massive fields, 
    \begin{equation}
     \label{Ccomponents}
C^{(\rm{CS})} = C  + \sum_{\rm massive~modes} \,. 
  \end{equation}   
   It is this massless $3$-form $C$ that 
  gives rise to   
   $\theta$-vacua. 
   
    For determining the    
 vacuum state, the massive modes are irrelevant, as their 
 contributions vanish in the zero momentum limit. 
 That is,  the massive modes (glueballs, or mesons) 
have zero occupation numbers in the vacuum state.
  This directly follows from the Poincar\'e invariance 
of the vacuum  \footnote{This statement should not be confused with a non-trivial contribution from an  infinite tower of 
 virtual massive modes into the correletor (\ref{EEcorr}).}.

  We thus reach a very important conclusion: 
 The vacuum of the $SU(N)$ theory 
 is fully determined by the massless 
 $3$-form $C$. 
 The corresponding vacuum equation is given by 
 (\ref{Evac}) and is solved 
 by an arbitrary constant $E_0$. This constant 
 represents an order parameter of $CP$-violation 
 by the vacuum of 
 QCD.  
 This is an exact statement. 
    
 Notice that we can think 
 of a constant $E_0 = \theta$ as of an 
 electric field of a massless $3$-form sourced by a boundary 
 ``brane" (\ref{BraneCoupling}) with the charge $Q = \theta$.  The existence of a massless pole in (\ref{CCcorr}) also makes it clear 
 how a boundary term  $\theta F\tilde{F}$
can have a local physical effect.

 In order to carry any information 
from the boundary to the local bulk, the existence of a  massless $3$-form field is necessary.  In the opposite case, the boundary term 
would have no physical effect,  since in a theory with only massive fields  at large distances all correlators die exponentially.

As already discussed, each solution 
$E_0 \neq 0$ represents a separate vacuum state with no 
possibility of a transition into others. 
  All of these vacua preserve the Poincar\'e invariance in form of 
  the space-time translations, rotations and Lorentz 
  boosts. However, the vacua with $E_0\neq 0$ break 
  $P$ and $CP$. It is clear that this breaking is spontaneous 
 since it comes from the VEV.

  The $E_0$-vacuum landscape is physically equivalent 
  to $\theta$-vacua.  
   In particular, the CP conserving vacuum 
  $\theta =0$ is the one with $E_0=0$. 
In pure glue,  $\theta \sim E_0$, in units of QCD scale,
which for convenience we have set equal to one.  
  
  Obviously, the language of the $3$-form \cite{Dvali:2005an} does not change the physics of $\theta$-vacua. However it provides an  understanding through a different prism. This prism relies solely on TSV.    
  This allows to better formulate the essence of 
     the strong-$CP$ puzzle and of the axion mechanism.   
   This formulation is also very convenient 
 for understanding the effect of gravity on strong $CP$.   
            
   In this language, the puzzle originates from the fact that among all possible 
   values of $E_0$,  offered by QCD, we happen to 
   live in the vacuum with an extraordinarily small  $E_0$ 
   ($E_0 \, \lesssim 10^{-9}$, in QCD units).  
  The theory with pure glue offers no proper explanation to this
  fact. 
  
  The situation changes upon the introduction of axion, 
  which renders the $3$-form massive.  The effect is in 
 Higgsing  the $3$-form.  At the level of  EFT,  axion 
 can be incorporated in two equivalent ways of $a$ and $B_{\mu\nu}$. 
  These are described by the Lagrangians (\ref{LaxionK}) and 
  (\ref{dual}) respectively.  The explicit nature of underlying
  physics that gives rise to these Lagrangians is unimportant for the 
  efficiency of the mechanism.  For example, in the original Peccei-Quinn realization, 
  the coupling between the axion and $E$ is generated through the fermion chiral anomaly.  However, it may as well originate at 
  more fundamental level, as shall be discussed later in the context
 of gravity. 

One way or the other,  the Higgsing of the $3$-form eliminates 
all vacua with non-zero $E$. 
  As a result, the true vacuum in QCD is the $CP$-conserving state 
 $E_0=0$. 
 \footnote{The $3$-form language of \cite{Dvali:2005an}
 clarifies the claim of \cite{Ai:2020ptm}
 that by changing the order of limits in ordinary instanton 
calculation, one ends up with $\theta=0$. In this approach one 
performs calculation in the finite volume and then takes it to infinity. 
In $3$-form language the meaning of this is rather transparent. 
   The finite volume is equivalent of introducing 
 an infrared cutoff in form of a shift of the massless pole 
 in (\ref{CCcorr}) away from zero. This effectively gives 
 a small mass to the $3$-form. For any non-zero value of the cutoff,
 the unique vacuum is $E_0=0$ which is equivalent 
 to $\theta =0$.  Other states  $E \neq 0$  (corresponding to $\theta \neq 0$) have finite lifetimes 
 which tend to infinity when cutoff is taken to zero. In this way 
 the $\theta \neq 0$ vacua are of course present 
 but one is constrained to 
 $\theta=0$ by the prescription of the calculation.
 Thus, changing the order of limits by no means eliminates 
 the $\theta$-vacua.  As usual, when taking the limit properly, one 
 must keep track of states that become stable in that limit.
 These are the states with $\theta \neq 0$ ($E \neq 0$), which become 
 the valid vacua in the infinite volume limit.  
 The effect is in certain sense equivalent to introducing an auxiliary axion and then decoupling it.}  

 We must note that the existence of the above vacuum structure 
 has an experimental verification from  
 the mass of the $\eta'$-meson.  This will be discussed in a separate section.   For now, we shall ignore the effect of the $\eta'$-meson.

 \subsubsection{$P$, $CP$ and Strong-$CP$}
 
 Although this is secondary to our main topic, 
 before moving to gravity, let us discuss 
 whether 
  the observed smallness of $\theta$ can be explained 
  by imposing a discrete 
  symmetry 
  on the Lagrangian.   Such approaches have been
  proposed in the literature \cite{Georgi:1978xz, Beg:1978mt, Mohapatra:1978fy, Nelson:1983zb, Barr:1984qx, Barr:1984qx}.
  The general idea is to 
  start with initial $\theta=0$ by imposing a symmetry,
  usually $P$ or $CP$.
  Of course, these symmetries are broken in the electroweak 
  sector. That breaking is assumed to be spontaneous, or at least soft. 
   The model then is constructed in such a way 
 that the dangerous contributions, such as 
  the phase of the 
 determinant of the quark mass matrix, are  under control and are small. 
 
  This approach does not take into account the fact that 
  $\theta$ itself breaks the ``protective" symmetry spontaneously.  
  As we have seen, the order parameter of this breaking is 
  the VEV of the ``electric" field $E_0$. 
  This VEV is arbitrary, regardless of the symmetry of the original 
  Lagrangian. 
  In this light, even if we put aside gravity, 
  the validity of the above solution is 
  questionable.   
  One can come up with the following arguments pro and contra. 
  
 {\it  Contra}: 
 
 The spontaneous breaking of a symmetry cannot 
 be restrained by the same symmetry, since a  
 solution of the theory need not respect the symmetry of the 
 Hamiltonian.  The physical $\theta$ is a solution that
 depends on an  integration 
 constant $E_0$. Due to this, the imposition of a symmetry, such 
 as $P$ or $CP$,  on the Lagrangian, 
    does not eliminate 
  vacua with arbitrary $\theta$, including $\theta \sim 1$. 
  Thus,  selecting $\theta=0$  has  no particular justification, 
  since other vacua are equally legitimate.

 {\it Pro} : 
  
 Due to absence of the transitions, 
 one may as well argue that a choice of the vacuum is equivalent 
 to a choice of the theory. One can therefore choose 
 $\theta=0$ by a symmetry.

 The above ambiguity is a peculiarity of the 
 superselection.      
  As we shall see, gravity eliminates it,
  leaving the axion as the only viable solution. 
  
  \subsection{Gravity} 
  
It has already been proposed that gravity necessitates the existence of axion \cite{Dvali:2018dce}.   The key argument was that, in the opposite
situation, the set of $\theta$-vacua would contain members 
with positive energy.  In a hypothetical world without gravity this is not an issue. The problem is gravity.  As argued previously 
\cite{us1},  such vacua are inconsistent with its quantum effects. 
From here, the conclusion was reached that 
gravity is incompatible with 
$\theta$-vacua.  
 Hence,  the presence of axion is a necessary condition for a consistent embedding of a gauge theory  in 
gravity.  
 Below, we shall support this conclusion and shall further elaborate on it. However, we shall rely on a 
 more general and powerful argument based on  the $S$-matrix
 \cite{Dvali:2020etd}.   
 
 The current understanding of quantum gravity 
  is fundamentally based on its 
$S$-matrix formulation. This puts the severe constraints on the 
vacuum landscape of the theory. 
In particular, it eliminates the possibility of de Sitter vacua, i.e., the vacua 
with positive energy densities.  However,  other cosmological 
space-times that do not asymptote to Minkowski vacuum,  are 
 equally problematic from the $S$-matrix perspective. 
 This list includes cosmologies with either collapsing or the eternally  inflating Universes. 

The detailed arguments can be found in  \cite{Dvali:2020etd}. Here, we shall 
take this knowledge as our starting point. 
 The necessity of the $S$-matrix vacuum has far-reaching consequences for the $\theta$-vacua 
 and the axion physics.  
   
  Let us first specify our framework.  
 In what follows, we shall be working within an effective low energy description of some unspecified 
fundamental theory of gravity. We shall 
assume that this fundamental theory is formulated through the $S$-matrix. 
  We shall restrict the asymptotic $S$-matrix vacuum to be 
Minkowski.  No other assumptions will be made. 
  
   This framework puts severe restrictions on the field content  of EFT.  
   In particular, it excludes the massless $3$-form fields. 
   The reason is very general. As already explained, 
   a massless $3$-form leads to the 
   existence of an infinite set of vacua. They are labeled by the VEV of the electric field  
  $E_0$. These vacua are non-degenerate in energy 
  and belong to distinct superselection sectors. 
    Due to this, a massless $3$-form  
produces     
 an  infinite number of superselection sectors 
  of non-Minkowski vacua.  Correspondingly, such a theory  
   does not satisfy our $S$-matrix criterion and cannot be embedded 
   in gravity. 
     
Thus, only the massive $3$-forms are compatible with
gravity.  This requires an axion. 
  In other words, the $S$-matrix formulation demands that every 
  $3$-form is accompanied by an axion which 
  Higgses it\footnote{The existence of a chiral fermion accomplishes the same goal, with axion being a composite of fermions.
  The example would be an $\eta'$-meson in QCD with a massless quark.}.  
  Obviously, this restriction extends to a 
 Chern-Simons $3$-form of any non-abelian gauge theory. 
 
    We  thus reach a surprisingly powerful conclusion.  When coupled to gravity, $\theta$-vacua must be eliminated.   
   That is, gravity demands the existence of axions, one 
   per each gauge sector with a non-trivial topological structure. 
   Among them, is of course QCD.

 \subsection{Necessity of exactness of axion mechanism}
    
   It is important to understand that the absence of $\theta$-vacua implies that the
 axion mechanism is exact.  In the opposite case,   
the unwanted superselection sectors cannot be avoided.   
  
   In other words,  the axion mass must come 
   strictly from its mixing with the corresponding 
   Chern-Simons $3$-form as in (\ref{LaxionK}). 
   Any additional explicit breaking of axion shift symmetry, 
   no matter how weak, is excluded.    
  As shown in \cite{Dvali:2017mpy},
  such an explicit breaking
 ``un-Higgses" the $3$-form and re-introduces the set of non-degenerate  $\theta$-vacua.  This violates the $S$-matrix 
  constraint. 
 
 It is easy to see this explicitly. 
  Let the mass of the axion be generated entirely from Higgsing the QCD $3$-form.
  This is described by the effective Lagrangian (\ref{LaxionK}).  
In this case, the vacuum is given by the $CP$-invariant 
state $E_0=0$. In the ordinary language, this 
implies that  $\bar{\theta}$ is unphysical.   

 Following \cite{Dvali:2017mpy}, let us now allow for some external contribution to the mass 
 of the axion.  We denote this increment by  $\mu^2$.  
 The corrected  Lagrangian 
  has the following form,   
   \begin{eqnarray} \label{Laxionmu}
 L &=&  \frac{1}{2} E^2 +  {1 \over 2} (\partial_{\mu} a)^2 -  {a \over f_a} E  -  {1 \over 2} \mu^2a^2\,,
\end{eqnarray}
where for simplicity we have taken ${\mathcal K}(E) =
\frac{1}{2} E^2$.  

The  equation of motion for the $3$-form is 
the same as in (\ref{EQE}) and is solved by, 
\begin{equation} \label{EAgr} 
E =  \frac{a- a_0}{f_a}\,.
\end{equation} 
As previously, $a_0$ is an integration constant which 
plays the role of the $\bar{\theta}$. 
In the absence of axion it would set the VEV of $CP$-odd 
field $E_0$. We can write $\bar{\theta} \equiv  \frac{a_0}{f_a}$.

 From (\ref{EAgr}) we get the following effective potential for the axion, 
 \begin{equation} \label{pot} 
 V(a) =  {1 \over 2} f_a^{-2} (a - a_0)^2 \, + \, {1 \over 2} \mu^2a^2 \,.      
  \end{equation} 
 
 Minimizing $V(a)$, we get 
  $a =  a_0\frac{1}{1 + \mu^2f_a^2}$.
  Plugging this into (\ref{EAgr}),
  we obtain the following expression for the VEV of the  $CP$-violating electric field, 
   \begin{equation} \label{pot} 
 E_0 \, = \, - \, 
   \frac{a_0}{f_a}\, \frac{\mu^2 f_a^{2}}{1 + \mu^2f_a^{2}}\, = \, -\,
   \bar{\theta} \, \frac{\mu^2 f_a^{2}}{1 + \mu^2f_a^{2}} \,.      
  \end{equation}
  We observe that for non-zero $\mu^2$, the VEV is proportional to the integration constant $a_0$.  
  
  In other words,  for $\mu \neq 0$, the $3$-form 
  field becomes un-Higgsed.  As a consequence,  the superselection sectors 
with different 
VEVs $E_0$,
re-emerge.   This VEV measures the amount of 
$CP$-violation. 
 Translated in the standard Peccei-Quinn 
language, for $\mu \neq 0$, the physical vacua with different values of  $\bar{\theta}$ re-emerge.

 At the same time, the energy densities in these vacua   
also depend on the integration constant, $a_0$, and correspondingly 
on $E_0$ or $\bar{\theta}$, 
\begin{equation} \label{poteff} 
 V_{min} \, = \, \frac{1}{2} \bar{\theta}^2 \, \frac{\mu^2f_a^2}{1 + \mu^2f_a^2} \,
 = \,  \frac{1}{2} E_0^2 \, \frac{1 + \mu^2f_a^2}{\mu^2f_a^2} \,.      
  \end{equation} 
  The lowest energy is 
  in the $CP$-conserving vacuum. There is no possibility to 
  have all the vacua in Minkowski.  
   Thus, such a theory cannot satisfy 
  our $S$-matrix constraint.

  The absence of non-degenerate vacua requires 
  $\mu=0$.  In this case,  only the $CP$-invariant 
  vacuum $E_0=0$ is present. 
  Thus, we see that gravity excludes any amount of 
  explicit breaking of axion symmetry.

  We are learning that not only gravity demands the 
 existence of axion per each $3$-form, it 
also requires sort of a ``fidelity" among the two. 
That is, the axion shift symmetry (\ref{axionshift}) must not 
be broken by any external source. 

  This is a rather curious result. 
  Ordinarily, gravity is assumed not to respect global symmetries.   
However, the situation we are encountering here is 
more radical.  
It appears that not only gravity respects the axion mechanism,  but also protects it from any external disturbance.

We  shall argue that this is an indication that gravity favors 
the formulation of QCD axion in terms of $B_{\mu\nu}$ field, described by the Lagrangian (\ref{LaxionK}) \cite{Dvali:2005an}. 
 In this formulation the protection 
is guaranteed by the QCD gauge redundancy. 
 This gives a protective power to the theory which is lacking 
in the standard Peccei-Quinn formulation. 
 Correspondingly, this also makes the theory predictive.   Let us explain this point
   carefully.

  \subsection{Gravity and the theory of axion}

 The EFT of axion (\ref{dual}) formulated in terms of 
 $B_{\mu\nu}$  is dual to 
  $a$-formulation given by (\ref{LaxionK}) 
  \cite{Dvali:2005an}. 
 However, these  EFTs are only valid below  the scale $f_a$.
 Above this scale, the axion requires an UV-completion. 
 In particular, both theories must properly merge with gravity. 
 In general, the UV-completions of $a$ and $B_{\mu\nu}$ 
 are different.  A clarification of this difference is important for
 understanding the influence of gravity on axion.  
   
  A generic EFT  that includes gravity is characterized 
 by a gravitational cutoff $M_{\rm gr}$.  This is the scale above which 
 the quantum gravitational effects become strong.
 In particular, gravity starts to dominate the scattering 
 processes  with the momentum-transfer per particle 
 exceeding $M_{\rm gr}$.   In general, the scale $M_{\rm gr}$ is lower  than the Planck mass, $M_P$.  For now, we shall keep it otherwise free.    
 
 Above the scale $M_{\rm gr}$, any EFT faces the 
 urgency of merging with quantum gravity.  This is regardless 
 whether other interactions are still weak at this scale. 
 Obviously, the need for an ultimate completion in gravity applies to both formulations of axion.  However, the two differ 
 by the possible forms of an intermediate completion.  
 
   If we  assume that $f_a <  M_{\rm gr}$, 
   then, above  $f_a$ and below  $M_{\rm gr}$,  
 the pseudo-scalar $a$ can be UV-completed 
 into a phase of a complex scalar field. 
 This is the case in the original Peccei-Quinn model. 
 However, in this completion the explicit breaking of axion symmetry can be achieved by a continuous deformation of the theory. 
   On the other hand, this would be incompatible with the $S$-matrix constraint of gravity.  Thus, such deformations must be absent in 
Peccei-Quinn. This must hold to all orders in field expansion. 
 However, EFT provides no visible reason for such a protection.  

In contrast,  the $B_{\mu\nu}$-formulation 
is protected from such 
deformations by the power of the gauge redundancy \cite{Dvali:2005an}.
  The renormalizable completion of $B_{\mu\nu}$ theory 
 below the scale $M_{\rm gr}$ is not known.   This implies that 
 in this theory  the axion constant $f_a$ cannot be below 
 $M_{\rm gr}$.  This is the good news for the theory, since it gives  a new correlation between the scales.  At the same time, there is no harm to calculability,  since below $f_a$ the theory 
 is weakly-interacting anyway. Quite the contrary, 
 the bonus is the 
 explicit protection of axion by the gauge redundancy already at the level of EFT.  This enables to predict $\bar{\theta} =0$.

  Due to above, we conclude that  gravity favors the 
$B_{\mu\nu}$-formulation of axion. 
Below, we consider the stability of the two formulations separately
and compare them.

 \subsection{UV-completion in Peccei-Quinn}  
 
 Let us assume that above the scale $f_a$ the axion $a$ is UV-completed 
  in form of a Nambu-Goldstone phase of a complex scalar field $\Phi = \rho(x) {\rm e}^{i\frac{a(x)}{f_a}}$,  where $ \rho(x)$ is the modulus. 
  The potential can be chosen as,  
 \begin{equation} \label{potPQ} 
 V(\Phi) = \lambda^2 (\Phi^{\dagger}\Phi -f_a^2)^2 \,. 
 \end{equation} 
  The axion scale $f_a$
 is set by the VEV 
  $\langle \rho \rangle = f_a$.
In this theory the axion shift symmetry (\ref{axionshift})
 is realized as a global  
 $U(1)_{PQ}$ transformation of the complex field,
 \begin{equation} \label{U1}
 \Phi \rightarrow {\rm e}^{i\alpha} \Phi\,,
 \end{equation} 
  by an arbitrary constant phase $\alpha$. 
 
 The gravitational consistency condition demands that the sole source of the explicit breaking
of $U(1)_{PQ}$ is the 
  coupling of axion with $E$, as it is given in the effective Lagrangian (\ref{LaxionK}). 
 In the Peccei-Quinn scenario, this coupling is generated through the fermion anomaly.   This is accomplished via coupling 
 $\Phi$ to $N_f$ flavors of quarks $\psi_j$  (with $j =1,2,...N_f$),
 which transform under the chiral symmetry (\ref{U1}) as 
 \begin{equation} \label{Upsi}
 \psi \rightarrow {\rm e}^{-i\frac{1}{2}\alpha \gamma_5} \psi\,.
 \end{equation} 
 In addition, the theory can include the set of $N'_f$ quark flavors, 
 $\psi'_r$  (with $r =1,2,...N'_f$), which do not transform 
 under the  $U(1)_{PQ}$-symmetry.   These quarks acquire their masses, $M'_r$, irrespective 
 of the spontaneous breaking of $U(1)_{PQ}$. 
 
   In different realizations of the Peccei-Quinn scenario
 \cite{Haxion1, Haxion2, Haxion3, Haxion4}, the 
ordinary and/or exotic quarks are assigned differently 
to the two sets.  This is a matter of the model-building which does not 
change the essence of the story.  The Lagrangian that captures the 
key physics of the Peccei-Quinn framework has the following form,  
   \begin{eqnarray} \label{LPQ}
 L_{\psi} \, =&& \, i \bar{\psi}_j \gamma^{\mu} D_{\mu} \psi_j  - g_j\Phi  \, \bar{\psi}_j  \psi_j \,  +  \nonumber \\
 && \, +  \, i \bar{\psi}'_r \gamma^{\mu} D_{\mu} \psi'_r \,  -\,  M'_r  \,\bar{\psi'}_r  \psi'_r  \, - \, \nonumber \\
  && \,  -  \,F_{\mu\nu}F^{\mu\nu}  - \, \theta \, F_{\mu\nu}\tilde{F}^{\mu\nu} \,,      
\end{eqnarray}
 where $D_{\mu}$ is the standard covariant derivative and 
 $g_j$ are the Yukawa coupling constants.
 The quantity $\theta$ is the ``initial" value of the boundary term
 which can be shifted by anomalous $U(1)_{PQ}$ transformation. 
  As previously, all irrelevant coefficients are set equal to one. 

 After taking into account the chiral fermion anomaly 
 and integrating over the instantons, the  
 theory reduces to the following EFT of the QCD $3$-form  and the axion, 
   \begin{eqnarray} \label{EcosA}
 L \, =&& \, E \arcsin(E) \,
 + \, \sqrt{1-E^2} \, +  {1 \over 2} (\partial_{\mu} a)^2 \, - \nonumber \\
 &&  - \,  \left ({a \over f_a} N_f - \bar{\theta} \right ) E \,,
\end{eqnarray}
 where, $\bar{\theta} \equiv \theta + {\rm arg.det.}(M')$. 
 We have chosen the form of the kinetic function
(\ref{Ecos}) which gives the $\cos$-type axion potential (\ref{Vcos})
\cite{Dvali:2005an}. 
In  the standard computations,  this form of the potential is obtained in the dilute instanton gas approximation.  As already discussed, the $3$-form picture 
shows why the mechanism works beyond this approximation, as 
the vacuum of the theory is 
totally insensitive to the form of ${\mathcal K}(E)$.  
 However, for the illustrative purposes, we shall work with  (\ref{Ecos}) in order 
to make the full parallel with the standard picture. 
Any other form of ${\mathcal K}(E)$, gives the exact same outcome. 

The equation of motion (\ref{EQE}) obtained from 
(\ref{EcosA}), 
\begin{equation}  \label{Esin} 
\partial_{\mu} \arcsin(E) = N_f\frac{\partial_{\mu} a}{f_a} \,, 
\end{equation} 
is solved by, 
\begin{equation}  \label{ENf} 
E = \sin \left (N_f\frac{a-a_0}{f_a} \right ) \,.
\end{equation} 
After plugging this into the equation  (\ref{EQE2}) for $a$, 
\begin{equation}  \label{ANf} 
\Box a + \frac{N_f}{f_a} E = 0 \,,
\end{equation} 
 we arrive to the following EFT of the 
axion, 
 \begin{eqnarray} \label{EFTPQ}
 L_a &=& {1 \over 2} (\partial_{\mu} a)^2 + 
  \cos\left (N_f\frac{a-a_0}{f_a} \right ) \,.
 \end{eqnarray}
 
 Notice that the entire information about the parameter 
 $\bar{\theta}$,  
  got absorbed into an over-all integration 
 constant, $a_0$,  
  \begin{eqnarray} \label{AzeroT} 
  N_f\frac{a_0}{f_a}  \equiv  \bar{\theta} \,.
 \end{eqnarray}
  The $3$-form language shows very clearly how the 
axion makes this quantity unphysical: The VEV of $E$ vanishes regardless of its value.   Of course, this is fully confirmed  by 
the minimization of the potential (\ref{EFTPQ}) which gives, 
 \begin{eqnarray} \label{AzeroT} 
  \bar{\theta}_{eff} = N_f\frac{a}{f_a} -  \bar{\theta} =0 \,.
 \end{eqnarray}
Obviously, the same outcome persists for a potential 
generated by an arbitrary form 
 of  ${\mathcal K}(E)$. \footnote{We comment that this result can be viewed as a generalization 
 of Vafa-Witten theorem \cite{VW}, which shows that 
 the global minimum of energy in QCD is $CP$-conserving. 
 The $3$-form language extends this conclusion to 
 every extremum of the axion potential, since 
 $E=0$ at each.}.

The potential in (\ref{EFTPQ})
respects the anomaly-free $Z_{N_f}$ subgroup of $U(1)_{PQ}$, 
under which the axion transforms as, 
\begin{equation} \label{ZN}
\frac{a}{f_a} \, \rightarrow \, \frac{a}{f_a} + l\frac{2\pi}{N_f} \,,
\end{equation} 
with integer $l$. 
Again, from (\ref{ANf}) and (\ref{ENf}) it is clear that, 
irrespective of the value of $a_0$,  the vacuum is $CP$-invariant.  In particular,  all Poincar\'e-invariant solutions 
give $E=0$.  

 Now, gravity tells us that any additional explicit breaking of 
 $U(1)_{PQ}$-symmetry is forbidden.
  For example, 
 all operators of the form,  
 \begin{equation} \label{NM} 
  (\Phi^{\dagger})^n(\Phi)^m \,, 
 \end{equation} 
 must vanish for $n\neq m$.   If present, in (\ref{EFTPQ}) they 
generate the additional terms
  of the form 
  \begin{equation} \label{PQNM} 
  V_{expl} \propto  \cos\left ((m-n)\frac{a}{f_a}\right )\,. 
 \end{equation}  
 As we have seen,  such explicit breaking produces a non-degenerate set of $CP$-violating 
vacua with $E_0 \neq 0$. 
 Obviously, this is incompatible with the $S$-matrix constraint. 
 Thus, due to the consistency requirement 
of gravity, all such $U(1)_{PQ}$-violating operators must vanish. 

The absence of the operators of the form  (\ref{NM}) goes against the standard expectation that such operators, 
suppressed by powers of $M_P$  (or $M_{\rm gr}$), 
can be included in EFT (see, e.g., \cite{Kamionkowski:1992mf, Holman:1992us, Barr:1992qq}).
In particular, wormholes as a potential source for 
breaking $U(1)_{PQ}$ were studied in \cite{Kallosh:1995hi} (for more discussion, see, \cite{Alonso:2017avz} 
and references therein).

 However, this is 
exactly what the $S$-matrix forbids.    
At the level of Peccei-Quinn theory of the $\Phi$-field,  this condition 
looks like a miracle.  
This theory possesses no obvious protection mechanism. Nor it exhibits any visible connection with gravity \footnote{ 
  Of course, one may try to 
 protect  $U(1)_{PQ}$ by  imposing a suitable discrete gauge symmetry. Such attempts have been 
made in the literature. However, in the present case 
this 
cannot 
work, since the protection must be exact. 
That is, the non-invariant operators must vanish to
all orders in field expansion. This is impossible to achieve by a finite discrete symmetry.}. 
 
  As show in \cite{Dvali:2005an},  the situation is 
  fundamentally different in $B_{\mu\nu}$-formulation of axion. 
  There, the axion mechanism is protected by the gauge invariance 
  to all orders in the operator expansion.  
   As already discussed, the $B_{\mu\nu}$ formulation can be 
  dualized to theory of $a$ with arbitrary potential
  $V(a-a_0)$ determined by the function ${\mathcal K}(E)$.
  However, the dualization from the theory of $B_{\mu\nu}$,
   always gives the theory of 
  $a$ with the integration constants, such as $a_0$. 
 Their number is controlled by the number of $3$-forms to which the $B_{\mu\nu}$ is coupled. 
  
  In contrast, the terms (\ref{PQNM}) 
 generated by explicit-breaking local operators (\ref{NM}) 
 in Peccei-Quinn model, do not possess such integration constants. 
 Correspondingly, such terms cannot be obtained by dualizing  
 the theory of $B_{\mu\nu}$.   This explains, why
 $B_{\mu\nu}$ formulation is more powerful  in protecting the
 axion mechanism, as compared to 
 the pseudo-Goldstone formulation. \\

  \subsection{Theory of gauged axion }
   
 Let us now discuss an alternative  theory
of the QCD axion formulated as the gauge theory of $B_{\mu\nu}$ field \cite{Dvali:2005an}. 
 We have already discussed the $B_{\mu\nu}$-formulation of axion 
 for a generic $3$-form $C_{\mu\nu\alpha}$.  The formulation for 
 the QCD axion is very similar,  with the role of $C$ played by 
 the massless component of the Chern-Simons $3$-form of gluons
 (\ref{CS}).  

 The idea is that, 
 at the level of a fundamental theory, 
   the QCD gauge fields are accompanied by a  
 Kalb-Ramond field $B_{\mu\nu}$. 
 This field shifts under the QCD gauge transformation as 
 (\ref{gaugeBC}).  
 This simple starting point fully fixes the outcome of the theory
 and determines its predictive power.

 Indeed, the same transformation acts on 
 $C^{\rm CS}$ as (\ref{gaugeC}). 
   Due to this, $B_{\mu\nu}$ enters the Lagrangian
     exclusively through a gauge invariant combination
   with the corresponding Chern-Simons $3$-form $C^{\rm CS}$,  
   \begin{eqnarray} \label{invC} 
 \bar{C}_{\alpha\beta\mu} \,  \equiv \, (C^{\rm CS} - f_adB)_{\alpha\beta\mu} \,,
\end{eqnarray}
 where $f_a$ is a scale.  
 Notice, the field $B$ is assigned a canonical dimensionality, whereas 
$C^{\rm CS}$ has the dimensionality $3$. 
Correspondingly, under the QCD gauge symmetry the two transform 
as
\begin{equation} \label{BQCD}
C^{\rm CS} \rightarrow C^{\rm CS} + d\Omega, ~~
B \rightarrow B + \frac{1}{f_a}\Omega,
\end{equation}
where, as previously, 
 $\Omega_{\mu\nu} = {\rm tr} A_{[\mu}\partial_{\nu]}\omega$. 
Of course, there also exists an additional redundancy 
(\ref{Bextra}) acting on $B_{\mu\nu}$.

 Thus, in the dual formulation, the axion emerges as an intrinsic 
 part of QCD that fully shares the gauge redundancy of gluons.  This is 
 the key to understanding the insensitivity of the QCD vacuum towards arbitrary local deformations of the theory. 
 
  Notice that we are making no assumption other than 
  adding a single degree of freedom  $B_{\mu\nu}$ to QCD. 
The gauge redundancy fixes the Lagrangian to be an arbitrary 
Poincar\'e invariant function of the gauge invariant quantities
$\bar{C}$ and $E \equiv \epsilon^{\nu\mu\alpha\beta}
\partial_{\nu}C^{\rm CS}_{\mu\alpha\beta}$.   

  That is, one can simply start at some cutoff scale 
 scale $M_{\rm gr}$  with the ordinary QCD Lagrangian, 
 $L_{QCD}$, with a single additional term fixed by the gauge symmetry (\ref{BQCD}), 
 \begin{eqnarray} \label{TheA} 
 L \, &=& \, L_{QCD}  \, + \, \bar{\theta} F\tilde{F}\, + \nonumber  \\
   &+&\, \frac{1}{f_a^2}(C^{\rm CS} \, - \, f_adB)^2 \,.
\end{eqnarray}
 With no further input, this is a fully accomplished theory of axion
 which makes $\bar{\theta}$ unphysical.  
 The mechanism is immune against the addition of arbitrary set of local 
 gauge invariant operators.  Correspondingly, it is stable 
 against UV quantum corrections, coming from gravity or any other physics.

As already discussed, the high derivative operators, 
as well as, all the massive modes contained in 
the spectral representation of the correlator (\ref{CCcorr}), can be set equal to zero without
the loss of any information about the vacuum structure. 
   Therefore, after integrating out all additional physics, the relevant part of the Lagrangian is fixed as,   
  \begin{eqnarray} \label{dualQ} 
 L &=&  {\mathcal K}(E) \, + \, \bar{\theta} E \, + \nonumber  \\
   &+&\, \frac{1}{f_a^2}(C^{\rm CS} \, - \, f_adB)^2 \,.
\end{eqnarray}
 The algebraic function ${\mathcal K}(E)$ is taken to be a generic function.  It is therefore free to accommodate all possible QCD and
 gravitational contributions that are algebraic in $E$. 
 
 However, in order to make a clear connection with the standard 
 language, we have singled out the linear term in $E$. 
  Of course, this is simply the standard boundary term,  
 $\bar{\theta} \,E \, = \, \bar{\theta} F\tilde{F}$, of  
the original theory (\ref{TheA}).  
  
 In addition, we can add arbitrary powers and/or the  mixed products of 
 $\bar{C}$ and $E$.  Due to the gauge invariance, they 
 leave the vacuum unchanged. 
  In short, the vacuum of the theory is insensitive to the 
 addition of arbitrary  local operators to (\ref{dualQ}).  
 
  The above theory makes it very clear why physical observables, such as EDMN, are independent 
 of  $\bar{\theta}$: 
  Due to the gauge-redundancy, the $3$-form is in the Higgs phase regardless the value of  $\bar{\theta}$. 
 Correspondingly,  all the $CP$-odd observables, such as the 
 VEV of  $E$,  vanish.   Let us make this more 
 explicit.

 As already explained, at low energies, the above theory 
reduces to an EFT of a would-be massless $3$-form $C$,  contained in the operator 
expansion of 
$C^{\rm CS}$, coupled to the axion $2$-form $B$. 
Therefore, in order to find the vacuum of theory, we must replace 
$C^{\rm CS} \rightarrow  C$.  

All Poincar\'e-invariant solutions of the 
  equations of motion,  
  \begin{eqnarray} \label{EQaxionB} 
\partial_{\mu}  \frac{\partial {\mathcal K}(E)}{\partial E}
  +   \frac{1}{f_a^2} \epsilon_{\mu\nu\beta\alpha}(C - f_adB)^{\nu\beta\alpha} &=& 0 \,,  \nonumber \\
  \partial^{\mu}  (C - f_adB)_{\mu\alpha\beta} & = &0  \,,      
\end{eqnarray}
give $E = 0$.
Correspondingly, the vacuum is exactly $CP$-conserving. 
  The gauge invariance guarantees that  QCD $3$-form is in the Higgs phase regardless 
  the form of the function ${\mathcal K}(E)$.   
  No gauge-invariant deformation of the theory 
can change this result. 

 For example, it is easy to check that 
the addition of arbitrary higher order terms in $\bar{C}$, 
leaves $E=0$ as the vacuum state.
 This is in accordance with the general arguments of \cite{Dvali:2005an} showing that in $B_{\mu\nu}$ formulation the axion
 vacuum is insensitive to arbitrary heavy physics. 
 That is, the $3$-form Higgs effect cannot be 
 abolished by an arbitrary set of massive modes.  
This is the key for explaining the axion protection ``miracle"
in $B_{\mu\nu}$-theory.    

One can explicitly check this statement by analysing the effect 
of arbitrary heavy sources on the $3$-form propagator
 \cite{Dvali:2005an} (see, \cite{Dvali:2005zk}, \cite{Sakhelashvili:2021eid}, for more recent analysis of explicit examples.) 
   
  As shown in \cite{Dvali:2005an}, the only gauge-invariant way of un-Higgsing the 
   QCD $3$-form, is the coupling of axion 
  $B_{\mu\nu}$  to an additional  massless $3$-form. 
   In  such a case,  one  linear superposition of the 
   two $3$-forms  remains  massless.  This would leave one physical family of  $CP$-violating $\theta$-vacua intact. 
   
    That is, in order to undo the axion mechanism,  
    we must change the theory discontinuously, via 
coupling the axion to an additional $3$-form $C'$. 
 In this way, the axion  $B_{\mu\nu}$ is ``shared" among the two 
$3$-forms,   
   \begin{eqnarray} \label{dualQ2} 
 L &=&  {\mathcal K}(E) + {\mathcal K}'(E') + \frac{1}{f_a^2}(C^{\rm CS} + C'- f_adB)^2 \,,
\end{eqnarray}
where $E'$ is the field strength of $C'$ and ${\mathcal K}'(E')$ is
its kinetic function.  Of course, $B_{\mu\nu}$ now 
transforms under the two independent gauge symmetries. 

As a result, only one linear superposition, $C^{\rm CS} + C'$,
of the two $3$-forms is Higgsed. The 
orthogonal superposition, $C^{\rm CS} - C'$, remains massless. 
Correspondingly, in such a theory 
the $\theta$-vacua are physical. 

By dualizing $B$ to $a$ and integrating out $C'$, 
 it is easy to see  that 
 we arrive to 
a theory of a pseudoscalar axion 
$a$ coupled to $E$, with the additional explicit breaking potential 
$V'(a - a_0')$.  This potential is determined from the equation (\ref{VK}) 
applied to ${\mathcal K}'(E')$. This gives, 
 \begin{equation}\label{KVaprime}
  \frac{\partial  {\mathcal K}'(E')}{\partial E'} 
   = \frac{a - a_0'}{f_a} \,.
 \end{equation}
 However, it is very important that the information about 
 the integrated-out $3$-form is encoded in the integration constant
 $a_0'$ \cite{Dvali:2005an}. 
 
 Instead, if one starts with the pseudo-Goldstone formulation of 
 Peccei-Quinn  axion with a fixed explicit-breaking potential (\ref{NM}),  the additional integration constant would be abolished. 
 This distinction is crucial for understanding 
 why,  unlike $B_{\mu\nu}$ case, there is no explicit  protection mechanism for the Peccei-Quinn formulation of the axion $a$.

 One of the consequences of the above is that  the 
 axion potentials with random explicit breaking terms 
 (\ref{PQNM}) cannot be dualized to $B_{\mu\nu}$ theory. 
   
 This shows a fundamental difference between the  
 $B_{\mu\nu}$-formulation and the formulation of axion in form of the phase of a complex scalar $\Phi$. 
  In the latter completion, the operators (\ref{NM}) that explicitly break  $U(1)_{PQ}$ symmetry, 
  are obtained by a continuous deformation 
 of the theory, without changing its field content.   
  In this light, the absence of such operators to all orders in $\Phi$-expansion looks mysterious.
    
 In contrast,  in 
 $B_{\mu\nu}$-formulation the axion mechanism 
 can only be jeopardised by the introduction of a new massless Chern-Simons $3$-form and the respective change of the gauge 
 properties of the axion. 
  However, this represents a discontinuous change of the theory
 already at the fundamental level.

    The bottomline is that the dual formulation makes 
   the protection of axion mechanism by gravity rather transparent. Gravity respects 
  the $3$-form Higgs effect of QCD due to gauge redundancy. 
  This fulfils the necessary condition for the $S$-matrix formulation of gravity.  

\subsection{Confronting the two theories of axion}

    Notice that no renormalizable 
 UV-completion in terms of a complex scalar field exist for the 
 $B$-formulation of axion. Instead, such a formulation likely requires the UV-completion directly in quantum theory of gravity. 
  Thus, for $B_{\mu\nu}$-formulation the gravitational origin of axion 
  is rather organic. 
  
   It is important to understand that the lack of known 
 renormalizable UV-completion for $B_{\mu\nu}$  cannot be taken 
 as a disadvantage as compared to the pseudo-scalar realization
 which can be embedded in the Peccei-Quinn model. 
   
   First, we must remember that the Peccei-Quinn theory is not UV-complete either.  It can certainly remain in weak-coupling regime up until the gravitational cutoff, but  the same is true about 
the $B_{\mu\nu}$-theory.  
  In addition, the question of embedding in gravity cannot be avoided  by either of the two theories.   
   
   Thus, the relevant question to be asked is by 
   how much the UV-completion into gravity (or other physics) impairs the predictive  power of the theory. 
 Here,  $B_{\mu\nu}$-theory (\ref{dualQ}) is at a clear advantage. Being equipped by the gauge redundancy (\ref{BQCD}), 
 its prediction $\bar{\theta}=0$ is respected by an 
 arbitrary UV-physics. 
 
     No such predictive power is exhibited by the standard 
  Peccei-Quinn realization which, with or without gravity, 
  is fully sensitive to the explicit-breaking operators of the type (\ref{NM}).  This makes it impossible to predict
the  value of $\bar{\theta}$.

\subsection{Harmless gravitational contribution to axion potential}

  As we showed, in $B_{\mu\nu}$-theory, the axion mechanism is 
 protected by the gauge symmetry to all orders in operator expansion. 
 By no means this implies that the gravitational contribution to the axion mass is zero. Rather, it only implies that  the gravitational  contribution is not un-Higgsing  the QCD $3$-form and 
 thereby respects the axion mechanism. 
 Nevertheless, this contribution can play an important role in phenomenology as we now explain. 
 
    The relevant corrections can come in two possible forms: 
   \begin{itemize}
  \item  The non-derivative
 functions of the invariants $E$ and $\bar{C}$; 
  \item A direct gravitational contribution into the TSV of QCD.
\end{itemize} 
  
 The first category effectively 
 reduces to a correction of the form of the kinetic function 
 ${\mathcal K}(E)$. Without loss of generality, we can split 
 this function as,  
   \begin{equation} \label{Ksplit}
  {\mathcal K}(E) \, = \,  {\mathcal K}_{QCD}(E) + 
  {\mathcal K}_{Gr}(E) \,, 
\end{equation}  
where ${\mathcal K}_{QCD}(E)$ accounts for pure-QCD   
 contributions, whereas ${\mathcal K}_{Gr}(E)$ 
accounts the gravitational effects.

The contribution  ${\mathcal K}_{Gr}(E)$ means that gravity 
generates some additional operators of $E$.
This contribution can be non-zero, even if a direct gravitational 
contribution into the TSV of QCD vanishes.  It is clear that 
${\mathcal K}_{Gr}(E)$ will correct the axion potential without 
jeopardising the $3$-form Higgs effect.
  The corrections can be estimated to the leading order in 
 field expansion.  
 
 For example, a gravitationally generated operator
 $E^2$ suppressed by 
 $M_P^4$,  generates a correction $\sim \Lambda^4/M_P^4$ into the kinetic 
 term of a canonically normalized  $3$-form. 
 This translates as a  
  relative correction to the axion mass of order $ \Lambda^4/M_P^4$,
 which is phenomenologically insignificant.  

 However,  potentially there exists a more important  
 gravitational correction to the axion mass in form of 
 the gravitational contribution into the TSV of QCD.
Under this we mean a gravitational contribution into the correlator 
(\ref{EEcorr}).
 This is important due to the fact that it corrects the spectral weight of  the massless pole in the K\"all\'en-Lehmann representation of the $3$-form correlator (\ref{CCcorr}).  This affects 
 the mass of the axion  as well as the mass of the $\eta'$-meson. 
 As we shall see, this contribution can be more substantial than
 the one coming from the high dimensional operators contributing 
 into (\ref{Ksplit}).  
   
 One potential source of gravitational contribution in TSV
  can come from the tower of  virtual black holes \cite{Dvali:2017mpy},  
   \begin{eqnarray}
     \label{EEBHcorr}
\langle E, E\rangle_{Gr} \, = \,
\sum_{BH}  \langle E \ket{BH} \bra{BH} E
\rangle  \, + \, ... \,.
\end{eqnarray}

  This expression shares some similarity with the contribution to TSV
  from the glueball tower in large-$N$ QCD in the framework of  the  Witten-Veneziano  mechanism \cite{Witten:1979vv, Veneziano:1979ec}.  
  In this context, Witten showed that the non-zero TSV in pure glue
  can be understood as a result of integrating out a 
  tower of glueballs. 
    
   Once we couple QCD to gravity, the black holes with all possible 
   glueball quantum numbers are present. 
  Since an off-shell glueball must have a 
  non-zero overlap with a corresponding black hole, 
  gravity is expected to produce the counterparts of the matrix 
  elements of the pure QCD. 
  
   We can estimate the semi-classical part of this contribution 
 using the arguments of  \cite{Dvali:2017mpy}. 
  A virtual black hole of entropy $S$ is expected to contribute 
  $\sim$e$^{-S}$.  Therefore, the dominant 
 contribution comes from the black holes of the 
 smallest entropy.  
 
  However, for the validity 
 of the semi-classical estimate, the black hole tower must be cutoff by the scale $M_{\rm gr}$ beyond which the semi-classical picture is not applicable. 
 In a theory with $N_{\rm sp}$ particle species, this cutoff is given by the species scale 
 \cite{Dvali:2007hz},  
 \begin{equation} \label{Species}
M_{\rm gr} = \frac{\sqrt{N_{\rm sp}}}{M_P}  \,.
 \end{equation}
 Correspondingly, the entropy of a cutoff-size black hole
 in Einstein gravity is $S = N_{\rm sp}$. 
 
Thus, in a theory with  $N_{\rm sp}$, the
 gravitational contribution to the TSV of QCD is expected to be exponentially suppressed as, 
 \begin{equation} \label{expN}
 \langle E, E\rangle_{Gr} \,
 \propto \,  {\rm e}^{- N_{\rm sp}} \,. 
 \end{equation} 
 Independently, there exist  a phenomenological bound 
 which comes from the mass of the $\eta'$-meson. 
 This mass would exceed 
 the experimental value,  if TSV would exceed the scale of QCD.  
 Thus, the gravitational contribution must be below the QCD scale. 
  The fact that the estimated contribution (\ref{expN}) 
it is exponentially small, goes in the right direction. 
  Beyond this, it is hard to put a constraint on  (\ref{EEBHcorr}). 
 
 Thus, at the current stage of our understanding, gravity could contribute 
 non-negligibly into the mass of the $\eta'$-meson. 
 It is hard to better bound this contribution due  
 to the fact that the QCD contribution into the TSV has not 
been computed fully.  

Through the same channel, gravity would simultaneously  contribute into the mass of the QCD axion. 
  Of course, none of the above contributions affect the Higgs mechanism of the QCD $3$-form. 
Correspondingly, they have no effect on exact $CP$-invariance  
of the QCD vacuum.   
 $\bar{\theta}$ remains unphysical.

\subsection{Gravitational TSV}  

One of the implications of the $S$-matrix criterion is that 
the TSV must vanish in each gauge sector of the theory.  
In particular,  this applies to the TSV of gravity.  This is the case
of special interest due to the universal nature of gravity.
 The correlator has the following form, 
  \begin{eqnarray}
   \label{RRcorr}
\langle R\tilde{R}, R\tilde{R}\rangle_{p\to 0} &&\equiv \\
\equiv \lim\limits_{p \to 0}\int d^4 x e^{ipx} \langle T [R\tilde{R}(0),
R\tilde{R}(x)]\rangle \nonumber\,, 
\end{eqnarray}
where $R$ is Riemann tensor and $\tilde{R}$ its dual. 
 The invariant $R\tilde{R}$, 
  \begin{equation} \label{dC}
  R\tilde{R} =  \epsilon^{\alpha\beta\mu\nu}
  \partial_{\alpha}C_{\beta\mu\nu}^{(\rm{G})}\,, 
   \end{equation}
 represents a field strength of the gravitational Chern-Simons $3$-form,  
  \begin{equation} \label{GCS}
   C_{\mu\nu\alpha}^{(\rm{G})} \equiv {\rm tr} (\Gamma_{[\mu}\partial_{\nu}\Gamma_{\alpha]} + {2 \over 3}\Gamma_{[\mu}\Gamma_{\nu}\Gamma_{\alpha]})\,,
   \end{equation}
   where $\Gamma_{\mu}$ is the connection and the contraction 
   of the suppressed indexes is obvious.   
   
 As explained in \cite{Dvali:2005an}, if the correlator 
 (\ref{RRcorr}) were a non-zero constant, this would imply 
 that  $C_{\mu\nu\alpha}^{(\rm{G})}$ contains a massless $3$-form. 
 The argument is identical to the one given in the case 
 of QCD and will not be repeated here. 
 
  As already discussed, such a $3$-form would produce the superselection sectors with different VEVs 
  of $R\tilde{R}$.  These vacua would have different energies and 
 would violate $CP$-symmetry  
  spontaneously.  
  They would serve as the gravitational analogs of 
  the QCD $\theta$-vacua. Their existence would conflict 
  with the $S$-matrix criterion.  
  Thus, the  necessary condition is that the gravitational TSV 
  is zero. 
  
  As discussed in \cite{Dvali:2017mpy}, one potential source contributing to the correlator (\ref{RRcorr}) is a tower of virtual black holes. The reasoning parallels the 
one given in the previous section 
 in the estimate of black hole contribution in the TSV of QCD.
Similarly to that estimate, the semi-classical black hole contribution to (\ref{RRcorr}) 
is expected to be exponentially suppressed by e$^{-N_{\rm sp}}$. 
Of course, there may exist other sources for the TSV of gravity.  
The important message is that due to the $S$-matrix constraint, 
the sum over all contributions must be zero.

 That is, if it happens that the contribution to (\ref{RRcorr}) 
from non-perturbative effects such as the virtual black holes
 is non-zero, this contribution must be exactly cancelled by some other physics. For this purpose,  gravity must be accompanied by a designated axion or a chiral fermion. 
  
   An interesting candidate is neutrino.  A neutrino with zero tree-level Yukawa coupling would make the gravitational $\theta$ unphysical due to the anomalous chiral 
  transformation \cite{Dvali:2013cpa, Dvali:2016uhn}. 
  This would be similar to the way a massless chiral quark makes $\theta$ unphysical in QCD.   
   In this process, the neutrino would get an effective mass, 
  similarly to the mass of $\eta'$-meson in QCD. 
  It has been speculated \cite{Dvali:2016uhn} that  this phenomenon can be the origin of small neutrino masses in the Standard Model. 

  The need for an additional axion, due to a mixed instanton contribution, 
 was also discussed in \cite{Chen:2021jcb}.  Again, this source  
 can be potentially problematic, provided it contributes to  (\ref{RRcorr}). In such a case it must be removed 
 either by an additional axion or by a chiral rotation of a would-be massless neutrino, as described in  \cite{Dvali:2013cpa, Dvali:2016uhn}.
   
  It also has been argued that gravity may contain a candidate for extra axion in form of torsion \cite{Karananas:2018nrj}.

  One way or the other, the $S$-matrix formulation
  demands that TSV of gravity is strictly zero.  This 
  is a prediction from quantum gravity.

\subsection{Gravitational origin of $f_a$}   
  
 From the fact that the presence of  axion is imposed by gravity,  
 we are naturally lead to the conclusion that $f_a$ is not independent of the scale of gravity. 
  This expectation is especially supported by the 
 $B_{\mu\nu}$-formulation of axion, which has no 
 renormalizable  UV-completion. As already discussed, in this formulation 
 $f_a$ is around or above the gravitational cutoff 
 $M_{\rm gr}$. On the other hand, the relation between $M_{\rm gr}$ and 
 $M_P$ is theory-dependent.  For example, as we know,   
 the particle species impose  an universal 
 upper bound on  $M_{\rm gr}$ given by (\ref{Species}). 
 
  Taking this as a 
  crude guideline, we can make some estimates. 
 In the minimal case, counting only the number of the Standard Model particle species plus graviton (around $120$), we put $f_a$ not far below the Planck mass.   
  Of course, at the level of the present discussion it is
  hard to come up with a precise relation.

  In this respect, two comments are in order.

 First,  $f_a$ can easily be as large as $M_P$
 without any conflict with cosmology. 
As shown in \cite{Dvali:1995ce}, the so-called standard cosmological upper bound, 
$f_a \lesssim 10^{12}$GeV \cite{Preskill:1982cy, Abbott:1982af, Dine:1982ah},  
is removed within the inflationary scenario due to
the fact that QCD could become strong in the early epoch. 
This would lead  
to an efficient dilution of the energy of the axion's  coherent oscillations. 
In this light, the axion can be cosmologically harmless, or even be 
a viable dark matter candidate, for $f_a$ as high as $\sim M_P$.

 Of course, having $f_a$ around the Planck scale 
 would make the direct searches of the QCD axion very 
 difficult. 
  However, in a generic theory, the scale $M_{\rm gr}$, 
 and therefore $f_a$, can be much stronger suppressed 
relative to $M_P$. 
  The important message however is that 
  $f_a$ cannot be below the cutoff of EFT.

 \subsection{Implications for non-axion approaches to strong-$CP$}  
 
 We briefly comment on the impact of our analysis on non-axion approaches to the strong-$CP$ problem. 
  Even though they may offer a way of selecting 
a vacuum with small $\theta$, the  
non-degenerate $\theta$-vacua of the full theory are 
still maintained.   As explained, this structure is incompatible with the $S$-matrix.  

  We can split non-axion approaches in two categories: 
  \begin{itemize}
  \item The models based on symmetries; 
  \item  The models based on cosmological 
  selection of the desired $\theta$-vacuum via dynamical and/or statistical mechanisms.  
\end{itemize}
  
 The earliest proposals from the first category are based on 
 $P$ or $CP$ symmetry
\cite{Georgi:1978xz, Beg:1978mt, Mohapatra:1978fy, Nelson:1983zb, Barr:1984qx, Barr:1991qx}.
 In these models the 
vacuum with small $\bar{\theta}$ is selected by imposing 
a discrete symmetry, $P$ or $CP$, 
on the initial Lagrangian. 
 
 As we discussed, already at non-gravitational level, the 
 solution is ambiguous, since $\bar{\theta}$ breaks
 parity and $CP$ spontaneously. However, gravity introduces a whole new dimension into the issue. Even if we somehow select  a 
 vacuum with favored value of  
$\bar{\theta}$, the other vacua are still part of the full theory. 
 Such a theory cannot be embedded in gravity due to the  $S$-matrix constraint.  The additional vacua must be eliminated and this requires 
 an axion. Correspondingly, the non-axion selection mechanism 
 becomes redundant. 

A  qualitatively different approach \cite{Dvali:2007iv}, 
is based on 
permutation symmetry between the large number of the hidden  copies of the Standard Model. 
This framework was introduced earlier \cite{Dvali:2007hz}
as the solution to the Hierarchy Problem, using the fact that a  
large number of hidden copies lowers the cutoff $M_{\rm gr}$ to the 
scale of species (\ref{Species}).  It was argued in \cite{Dvali:2007iv} that,  as a byproduct, this setup simultaneously 
lowers the value of $\bar{\theta}$.

Notice that the permutation symmetry imposed on the 
initial Lagrangian does not imply 
the equality of the $\bar{\theta}$-terms in different copies of QCD. 
This is because - just 
as they do with $P$ and $CP$ - these terms 
break the permutation symmetry spontaneously. 
However,  the sum of the $\bar{\theta}$-parameters over all sectors 
is bounded from above by the universal cutoff (\ref{Species}). 
It was then shown that statistically most probable value
of the $\bar{\theta}$-term comes out to be close to its phenomenological bound (\ref{BBartheta}). 
Nevertheless, this approach does not eliminate the non-degenerate 
$\theta$-vacua 
fully. It is therefore incompatible with the $S$-matrix.  

The second category of non-axion models, 
 is based on a dynamical cosmological selection of the 
 $\theta$-term.   To our knowledge, the earliest representative 
 of this idea is \cite{Dvali:2005zk}.
 This model employs a so-called ``attractor"  
 mechanism, originally introduced in  
 \cite{Dvali:2003br, Dvali:2004tma} for the 
 cosmological relaxation of the Higgs mass. 
 Although, in the  scenario of \cite{Dvali:2005zk} $\bar{\theta}$ is 
 a dynamical field, unlike axion, its evolution is 
 dominated by heavy physics.  However,  this physics is
back-controlled by $\bar{\theta}$ in the way that  makes 
 $\bar{\theta} \rightarrow 0$ into an attractor of the cosmological evolution. 
 Unfortunately, this scenario does not eliminate vacua with other values $\bar{\theta}$. Many of these are long lived 
 (eternally inflating) de Sitter vacua.  Due to this, they are in conflict with the $S$-matrix constraint.  
 
 Another representative  of the cosmological selection  is the 
 scenario of \cite{TitoDAgnolo:2021nhd}. This model is free from 
 the presence of the eternal de Sitter vacua. However, 
 instead it incorporates  the cosmological branches with asymptotically-collapsing anti de Sitter space-times. These too are
incompatible with the $S$-matrix.

 We must comment that despite the above difficulties, the issue with cosmological relaxation is not closed.  One could envisage 
 some modification of the scenarios of the type 
 \cite{Dvali:2005zk} or \cite{TitoDAgnolo:2021nhd} in which 
 the unwanted vacua would only exist temporary and would not affect 
 the asymptotic future.  This can in principle be reconciled 
 with the $S$-matrix. 
   An interesting question is whether such ``refined" scenarios, 
 effectively become equivalent to an ordinary axion 
  solution.   
  
  There exist number of other interesting representatives 
  from both categories, which are not reviewed here 
 due to lack of space.  The important thing is to draw a general 
 lesson: The approaches that somehow select 
 a desired vacuum with a specific value of $\bar{\theta}$, while maintaining the vacua with 
 other values of $\bar{\theta}$, are problematic for coupling to gravity.  
  
 \subsection{Role  of $\eta'$} 
 
 We wish to touch upon the role of the $\eta'$-meson. 
Let us imagine a situation in which at least one of the quarks, in the Standard Model possesses an axial symmetry (\ref{Upsi}), exclusively broken by the QCD anomaly.   Of course, this would require   
a strictly zero Yukawa coupling with the Higgs field,
as well as, the absence of all non-invariant higher dimensional operators.  This appears to be excluded phenomenologically \cite{Leutwyler:2000hx} (for more recent analysis, see,
\cite{Alexandrou:2020bkd}, and references therein). 
Nevertheless, the regime of massless chiral quarks is very useful for the theoretical understanding.  
 
In such a situation there would be no need for the axion. 
Or, to be more precise, the role of the axion would be taken-up 
by the $\eta'$-meson \cite{Dvali:2005an}. 
The axial symmetry of the massless quark would play the role of 
$U(1)_{PQ}$-symmetry.   This symmetry is spontaneously broken by the quark condensate. 

 As 't Hooft showed \cite{tHooft:1986ooh}, the corresponding would-be Goldstone boson, $\eta'$ gets a mass from instantons through 
 its anomalous coupling with $F\tilde{F}$.  
 As it is well-known, this solves the $U(1)$-problem. However, 
 if one of the quarks were massless, simultaneously, 
 the $\eta'$ would solve the strong-$CP$ problem,  exactly the way this is done by the axion.
    
 The generation of $\eta'$ mass can also be understood in terms 
 of TSV \cite{Witten:1979vv, Veneziano:1979ec}, without 
 an explicit reference to instantons. 
 
   In the real world description of the Standard Model, the axial symmetry of quarks is explicitly broken by their Yukawa couplings. 
   Correspondingly,  $\eta'$ cannot  eliminate $\theta$-vacua and
   axion is necessary. However, it still plays an important role for our 
   analysis, as it represents an experimental proof of the existence 
   of the $\theta$-vacua. 
   
    This fact is very important for eliminating a potential loophole 
    in our statement that axion is necessary for 
    the $S$-matrix formulation of gravity.  Without 
  the $\eta'$-meson, the following loophole 
    can be considered \footnote{We thanks 
    Otari Sakhelashvili for raising this question.}. 
       Let us restrict the Hilbert space of the theory  
  to a particular $\theta$-vacuum which 
    satisfies the $S$-matrix criterion.  At the same time, we exclude all others. 
  In other words, we imagine that, instead of explaining their absence 
 dynamically, via axion relaxation, gravity instructs us to 
 simply forget all values of $\theta$ that do not 
 correspond to Minkowski vacua.  
 
 The mass of the $\eta'$ is the evidence that gravity did not take this path.  It represents a proof of the existence of the $\theta$-vacua
 as part of the theory.
 The coherent excitations of the $\eta'$-field, are nothing but the 
 excursions into the neighbouring $\theta$-vacua. 
 
 Notice that $\eta'$ somewhat reduces the amount of 
 $CP$-violation triggered by the $\theta$-vacua.  
 In the language of the $3$-form description, the effect of 
 $\eta'$ can be understood as a ``partial-Higgsing".  
 This is accounted by the effective Lagrangian of the 
 type (\ref{Laxionmu}) in which $a$ is replaced by 
 $\eta'$,  
    \begin{eqnarray} \label{Letamu}
 L &=&  \frac{1}{2} E^2 +  {1 \over 2} (\partial_{\mu} \eta')^2 -  {\eta' \over f_{\eta}} E  -  {1 \over 2} \mu^2\eta^{'2}\,.
\end{eqnarray}
The decay constant is of order the QCD scale $f_{\eta} \sim 1$.  
The quantity $\mu^2$ accounts for explicit breaking of chiral symmetry by non-zero quark masses.  It  vanishes
in the chiral limit when at least  one of the quarks is taken massless.  Since some quarks are lighter that the QCD scale, 
 we have $\mu^2 \ll f_{\eta}$. 

  Integrating out the $3$-form through its equation of motion,
  we get, 
     \begin{equation} \label{Eeta} 
 E \, = \, \frac{\eta'}{f_{\eta}} \,-\, \bar{\theta} \,,       
  \end{equation}
where $\bar{\theta}$ is an integration constant.
Plugging this into the equation for $\eta'$, we obtain the following
expressions for its VEV, 
   \begin{equation} \label{Veta} 
\eta' \, = \, \bar{\theta} \frac{f_{\eta}}{1 + \mu^2 f_{\eta}^2}\,,
 \end{equation}
 and the mass, 
  \begin{equation} \label{Meta} 
 m^2_{\eta} \, = \, f_{\eta}^{-2} + \mu^2 \,.
 \end{equation}
 It is clear that the contribution of the 
 explicit breaking-term is 
 subdominant as compared to the contribution from 
 the $3$-form.   

 Due to this, the contribution from $\bar{\theta}$ into the electric field 
 is partially cancelled. This can be easily seen by 
 plugging (\ref{Veta}) into (\ref{Eeta}), which gives, 
   \begin{equation} \label{Eeta1} 
 E_0 \, = \,-\, \bar{\theta}
   \frac{\mu^2f_{\eta}^{2}}{1 + \mu^2 f_{\eta}^{2}} \,.       
  \end{equation} 
  It is clear from  both (\ref{Eeta}) and (\ref{Eeta1}) 
  that in the limit of a chiral quark, 
 $\mu^2 =0$,  the $\eta'$ fully eliminates 
 the dependence on $\bar{\theta}$ and we end up with an unique 
 $CP$-invariant vacuum, $E=0$.
  In this limit $\eta'$ becomes an axion.  

 We would like to remark that the above generation of the mass of $\eta'$ 
can also be understood in purely topological terms 
\cite{Dvali:2005ws}.   
    
 In the real world, due to non-zero quark masses,  the $\eta'$-meson cannot fully eliminate the $CP$-odd electric field $E$.
 Correspondingly, the $\theta$-vacua persist
 and $\bar{\theta}$ is physical. 
  As a result, the $S$-matrix criterion requires an axion.

 \subsection{EDM of neutron}   
 
  Within the Standard Model, the QCD contribution to the 
  electric dipole moment of neutron (EDMN)
  is controlled by the quantity $\bar{\theta}$ \cite{Baluni:1978rf, Crewther:1979pi}. 
  The role of this parameter in $3$-form formulation
   is taken up by the integration constant, $a_0/f_a \equiv \bar{\theta}$.  The axion renders this quantity unphysical, 
  provided the axion shift symmetry (\ref{axionshift}) is 
 broken exclusively by the QCD anomaly. 
 In such a case the QCD contribution to EDMN vanishes.
 
 Within the Standard Model, there exists an additional contribution 
  to EDMN coming from the electroweak breaking 
  of $CP$-invariance \cite{Ellis:1976fn, Shabalin:1979gh, Ellis:1978hq}. 
   However, the predicted value, 
  $|d_n| <  2.9 \times 10^{-32} e$cm,  is 
  unmeasurably  small.  
 It is minuscule as compared to the current experimental 
 limit  $|d_n| <  2.9 \times 10^{-26} e$cm \cite{Baker:2006ts}. 
It is also hopeless to be detected in   
a foreseeable future (e.g., see \cite{n2EDM:2021yah}). 
    
  However, within the standard Peccei-Quinn 
 approach the above cannot be taken as a prediction, since the global 
  $U(1)_{PQ}$-symmetry can be 
  explicitly broken, arbitrarily.  This introduces an 
  uncalculable $\bar{\theta}$-dependent contribution to EDMN.  
  
  We have argued that gravity makes such a breaking impossible.  
 That is, in gravity, the axion mechanism must remain fully undisturbed.   Correspondingly,  $\bar{\theta}$ must be strictly unphysical. 

   Somewhat surprisingly, this emerges as an absolute statement. 
   However, it is a direct consequence of a more fundamental 
  absolute statement:  
   The exactness 
    of the axion mechanism      
  imposed by gravity.     
  We have explained that this exactness can be 
 understood as a consequence of the QCD gauge redundancy
 (\ref{gaugeBC}). This is intrinsic in the  
gauge formulation of the axion mechanism
described as 
the $3$-form Higgs effect \cite{Dvali:2005an}. 
In this theory, the independence of physics on $\bar{\theta}$ is a prediction. 

  This implies that 
 there is no contribution to EDMN from $\bar{\theta}$. 
  The phenomenological lesson from here 
  is that any experimental indication of non-zero EDMN 
 will be a probe of new physics beyond the Standard Model. 
  In particular, such evidence will be explainable neither in terms of the effective $\bar{\theta}$
 nor in terms of the weak $CP$-violation. 
  The former is strictly zero, whereas the latter  \cite{Shabalin:1979gh, Ellis:1978hq}  is
 by four orders of magnitude below the prospective 
 experimental sensitivity \cite{n2EDM:2021yah}. 
   Due to this, the non-zero signature of EDMN can only come from
   new physics.

 The bottomline is that EDMN represents an unambiguous experimental probe  
 of new physics.  In this light,  it is important to analyse the contributions to EDMN from the motivated  extensions of the 
 Standard Model.  Such analysis can be performed \cite{OtoGia} directly
within the EFT of chiral Lagrangian in the spirit of \cite{Bsaisou:2014oka, Bsaisou:2014zwa}.

   For a reader less interested in gravity, the results of this chapter 
   can still be useful \footnote{We are grateful to Goran Senjanovic
   who for a long time was insisting on the need of such a clarification. 
   We thank  Otari Sakhelashvili  and Akaki Rusetsky for valuable discussions.}.  It addresses the  question of EDMN 
  in the presence of exact axion mechanism, irrespective of its origin. 
  The arguments indicating that 
  \begin{equation}
  {\rm Detection~of~ EDMN \, = \,New ~Physics} \,, 
  \end{equation}
  are insensitive to what guarantees the protection of the 
  axion  mechanism.

  \subsection{Summary and outlook}
  
  The present paper brings the two main points. 
 First, gravity demands the existence of an ``undisturbed" axion
 per each gauge sector.  Secondly, due to this, gravity  
 favors the 
 theory of the QCD axion based exclusively on the gauge 
 redundancy of gluons, without involvement of a 
 global $U(1)_{PQ}$-symmetry. 
  
  This solution to strong-$CP$ puzzle was studied earlier in \cite{Dvali:2005an}, where it was shown to be stable 
 against  deformations by arbitrary local operators. It therefore emerges as a viable setup that reconciles the two seemingly-conflicting tendencies of gravity: \\
   
   1)  The absence of $\theta$-vacua, required  for the consistency of 
   the $S$-matrix formulation;  
   
   and

   2) no respect towards the global symmetries.  \\
   
    The first requirement demands the presence of axions with exact 
 relaxation mechanisms.   This is difficult to achieve in the standard 
 Peccei-Quinn setup, since the global $U(1)_{PQ}$-symmetry
 need not be respected by gravity.  
 
   In contrast, as shown in \cite{Dvali:2005an} and discussed at length here,  the $B_{\mu\nu}$-theory, which does not rely on any global symmetry,  is fully immune due to the gauge redundancy.   
  In this theory the axion relaxation is substituted 
  by the Higgs-like effect in which the QCD Chern-Simons 
  $3$-form  becomes massive by eating-up the $B_{\mu\nu}$-axion. 
  This relaxes the $\theta$-vacua to the ground state
  in which all the $CP$-odd observables vanish.    
     Stability of the $3$-form Higgs phase is guaranteed by the gauge redundancy of QCD.  This ensures the exactness of the mechanism  to all orders in operator expansion.

   Besides, the gauge theory of $B_{\mu\nu}$ has number of advantages over the standard approach.  
   First, it is remarkably simple.  All one needs to do, is 
   to introduce a single degree of freedom, $B_{\mu\nu}$, 
   with a proper gauge charge under QCD.  
   
    With no further assumptions, the theory takes care of itself. 
 Due to the gauge redundancy, the structure 
  and the outcome of the theory are determined unambiguously. 
  
   Secondly, the theory has a predictive power, implying 
   $\bar{\theta}=0$.   
    This goes in contrast with the standard 
   Peccei-Quinn setup in which $U(1)_{PQ}$ is not protected by 
   any fundamental principle. Correspondingly, 
   there $\bar{\theta}$ is not calculable. 
   
   Due to the 
exactness of the axion mechanism, an important  phenomenological prediction is that the  QCD contribution to EDMN, 
which is set by $\bar{\theta}$, is exactly zero. 
At the same time,  the contribution from the electroweak $CP$-violation is by six orders of magnitude below the current experimental limit \cite{Baker:2006ts} 
and by four orders magnitude below the aim of the planned  experiments (see, e.g., \cite{n2EDM:2021yah}). 
Due to this, unless the precision is improved substantially,
the electroweak contribution to EDMN cannot be measured. 
Within such sensitivity, any detection of EDMN will be a signal of
  a new $CP$-violating physics, coming from beyond QCD and weak interactions.

   There are number of other implications. 
   Through its $S$-matrix formulation, quantum gravity predicts that 
   TSV must vanish in each sector of the theory.  In particular, 
   the TSV of gravity must also be zero. In this light, it is important to understand whether some non-perturbative entities, such as virtual black holes \cite{Dvali:2017mpy}, contribute non-trivially to this correlator.   
 If the answer  is positive, this would predict the existence 
 of a new degree of freedom - the second (gravitational) axion or a chiral    
  fermion - designated for an exact cancellation of such contributions into the gravitational TSV \cite{Dvali:2005an}. 
  
  The standard model contains an interesting candidate in form of 
  neutrino \cite{Dvali:2013cpa, Dvali:2016uhn}. 
 In the absence of a tree-level Yukawa couplings for one 
 of the neutrino species, the gravitational TSV would identically  vanish due to the gravitational anomaly in the chiral neutrino 
 current.  \\

{\textsl{\bf Acknowledgments}}\;---\;We thank  Akaki Rusetsky, Otari Sakhelashvili and Goran Senjanovi\'c
for discussions. 
This work was supported in part by the Humboldt Foundation under Humboldt Professorship Award, by the Deutsche Forschungsgemeinschaft (DFG, German Research Foundation) under Germany's Excellence Strategy - EXC-2111 - 390814868, and Germany's Excellence Strategy under Excellence Cluster Origins.

\end{document}